\documentclass[a4paper,fleqn,usenatbib]{mnras}

\usepackage{mathptmx}

\usepackage[T1]{fontenc}
\usepackage{ae,aecompl}

\usepackage{graphicx}	
\usepackage{amsmath}	
\usepackage{amssymb}	
\usepackage{epsfig}
\usepackage{hyperref}
\usepackage[usenames, dvipsnames]{color}

\usepackage{times}

\title[Molecular gas properties of Virgo spirals]{The effect of ram pressure on the molecular gas of galaxies: three case studies in the Virgo cluster}

\author[B. Lee et al.]
{Bumhyun Lee$^{1}$\thanks{E-mail : bhlee301@gmail.com}, Aeree Chung$^{1,2,3}$\thanks{E-mail : achung@yonsei.ac.kr (corresponding)}, Stephanie Tonnesen$^{4}$,  Jeffrey~D.~P. Kenney$^{5}$, \and O. Ivy Wong$^{6}$, B. Vollmer$^{7}$, Glen R. Petitpas$^{8}$, Hugh~H. Crowl$^{9}$, Jacqueline van Gorkom$^{10}$
\\
\\
$^{1}$Department of Astronomy, Yonsei University, 50 Yonsei-ro, Seodaemun-gu, Seoul 03722, Korea \\
$^{2}$Yonsei University Observatory, Yonsei University, 50 Yonsei-ro, Seodaemun-gu, Seoul 03722, Korea \\
$^{3}$Joint ALMA Observatory, Alonso de C{\'o}rdova 3107 Vitacura, Santiago, Chile \\
$^{4}$Carnegie Observatories, 813 Santa Barbara St, Pasadena, CA, 91101 \\
$^{5}$Yale University Astronomy Department, PO Box 208101, New Haven, CT 06520-8101, USA \\
$^{6}$International Centre for Radio Astronomy Research, The University of Western Australia M468, 35 Stirling Highway, Crawley, WA 6009, Australia \\
$^{7}$CDS, Observatoire astronomique de Strasbourg, Universit\'e de Strasbourg, CNRS, UMR 7550, 11 rue de l'Universit\'e, F-67000 Strasbourg, France\\
$^{8}$Harvard-Smithsonian Center for Astrophysics, 60 Garden Street, Cambridge, MA 02138, USA \\
$^{9}$Division of Science and Mathematics, Bennington College, 1 College Drive, Bennington, VT 05201, USA \\
$^{10}$Department of Astronomy, Columbia University, Mail Code 5246, 550 W 120th St, New York, NY 10027, USA \\
}

\date{Accepted to MNRAS}

\pubyear{2016}

\begin{document}
\label{firstpage}
\pagerange{\pageref{firstpage}--\pageref{lastpage}}
\maketitle

\begin{abstract}
We present 12CO (2--1) data of three Virgo spirals -- NGC~4330, NGC~4402 and NGC~4522 obtained using the Submillimeter Array. These three galaxies show clear evidence of ram pressure stripping due to the cluster medium as found in previous H{\sc i} imaging studies. Using high-resolution CO data, we investigate how the properties of the inner molecular gas disc change while a galaxy is undergoing H{\sc i} stripping in the cluster. At given sensitivity limits, we do not find any clear signs of molecular gas stripping. However, both its morphology and kinematics appear to be quite disturbed as those of H{\sc i}. Morphological peculiarities present in the molecular and atomic gas are closely related with each other, suggesting that molecular gas can be also affected by strong ICM pressure even if it is not stripped. CO is found to be modestly enhanced along the upstream sides in these galaxies, which may change the local star formation activity in the disc. Indeed, the distribution of H$\alpha$ emission, a tracer of recent star formation, well coincides with that of the molecular gas, revealing enhancements near the local CO peak or along the CO compression. FUV and H$\alpha$ share some properties in common, but FUV is always more extended than CO/H$\alpha$ in the three galaxies, implying that the star-forming disc is rapidly shrinking as the molecular gas properties have changed. We discuss how ICM pressure affects dense molecular gas and hence star formation properties while diffuse atomic gas is being removed from a galaxy.
\end{abstract}

\begin{keywords}
ISM: molecules -- galaxies: clusters: intracluster medium -- galaxies: evolution -- galaxies: ISM -- galaxies: spiral.
\end{keywords}



\section{INTRODUCTION}

Since \cite{gunn1972} suggested that a galaxy might lose its interstellar medium (ISM) by interacting with the intracluster medium (ICM) in the cluster environment, much evidence for ram pressure stripping has been found to date. In early days, \cite{davies1973} have shown that Virgo galaxies are generally more deficient in H{\sc i} compared to field galaxies based on their single-dish observations. In addition, \cite{giovanelli1985} have found that the H{\sc i} content is well correlated with the location of galaxies in a sense that H{\sc i} is more deficient at smaller distances from the cluster centre. Then the H{\sc i} synthesis imaging studies such as \cite{warmels1988a,warmels1988b} and \cite{cayatte1990} have shown that H{\sc i}-deficient galaxies near the cluster centre have small H{\sc i}-to-optical extents, truncated within the stellar disc in many cases. In a more recent high-resolution H{\sc i} imaging study of $\sim$50 selected Virgo galaxies by \cite{chung2009}, a number of galaxies have been found with various scales of extraplanar H{\sc i} gas or long H{\sc i} tails, indicating that ram pressure stripping is indeed acting in the cluster environment.

As a galaxy loses its ISM, the star formation rate is expected to be suppressed, which agrees well with the observations. As a good example, \cite{koopmann2004a} find that the massive star formation rate of Virgo spirals is lower than their field counterpart by a factor of 2.5 on average. \cite{koopmann2004b} also show that many Virgo spirals have a smaller H$\alpha$ extent compared to their stellar disc, reflecting that the process truncating  H{\sc i} discs may also be acting on the star-forming disc and H$\alpha$ emission.

On the other hand, molecular gas is unlikely to be as easily stripped as atomic hydrogen, since it is more tightly bound to the galactic centre and the density is higher. In fact, most previous studies find that the molecular gas mass of the cluster population is not significantly different from that of field galaxies \citep{stark1986, kenney1989}. In addition, more recent studies are finding clumpy dust features in the upstream side of H{\sc i} gas stripped galaxies, which are likely to be surviving dense clouds that are unveiled after diffuse atomic gas is removed (\citealt{crowl2005,abramson2014}; Kenney, Abramson \& Bravo-Alfaro 2015).

On the other hand, however, the opposite results also have been reported. \cite{rengarajan1992} find that H$_{2}$ mass normalized by the dynamical mass of galaxies tends to get larger with increasing clustercentric distance, which supports that molecular gas can be also deficient in the cluster environment. More recently, \cite{boselli2014} show that H{\sc i}-deficient galaxies in the cluster environment tend to be also modestly deficient in molecular gas.

However, the molecular gas fraction to the optical luminosity of spiral galaxies measured using 12CO (1$-$0) ranges quite widely for all environments \citep{chung2012}. This implies that the molecular gas fraction alone may not be a good tracer of molecular gas deficiency. Due to the scatter in this relation and opposing observational results, it is still arguable whether molecular gas can be affected by the ICM pressure in a similar way to H{\sc i} gas, and it is then still puzzling why star formation appears to be quenched in H{\sc i} stripped galaxies if molecular gas, which is the more direct ingredient for star formation, is not deficient.

\begin{figure}
\begin{center}
\includegraphics[scale=0.41,bb=10 70 585 770]{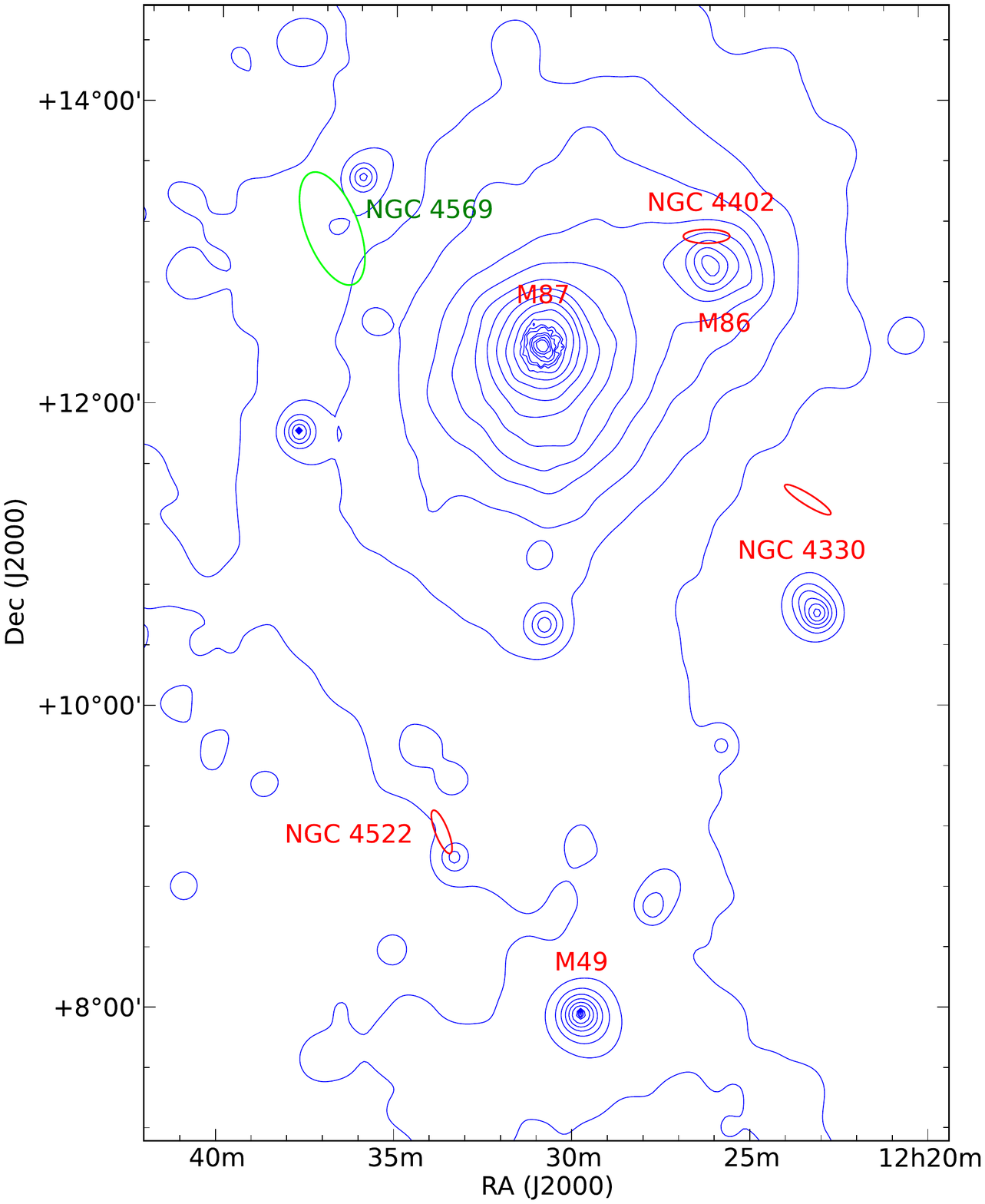}
\end{center}
\caption[]{The locations of our sample are shown on the {\it ROSAT} X-ray map (blue contours; \citealt{bohringer1994}). Red ellipses represent the position angle and $D_{25}\times5$ of the sample in the {\it B} band. NGC~4569 that we have not observed using the SMA but is included in our discussion as a good representative of a galaxy at post-peak pressure, is shown in green. The sample galaxies, located at 0.4--1 Mpc from M87, adopting a Virgo distance of 16 Mpc \citep{yasuda1997}, make a nice sequence of ram pressure stripping from early stage, close to peak pressure and post-peak pressure in their orbit.} 
\label{vir_pos}
\end{figure}

No clear evidence for molecular gas stripping yet low star formation activities in cluster galaxies may imply distinct molecular gas properties in high-density environments as supported by some previous observations. For example, \cite{kenney1990} have found a highly asymmetric CO morphology in an H{\sc i} deficient Virgo cluster galaxy. \cite{vollmer2008} also have found very peculiar CO distributions in some of Virgo spirals undergoing active H{\sc i} stripping. These studies indicate that molecular gas can be potentially disturbed by strong ICM pressure whether it is stripped or not.

Particularly along the side experiencing ICM pressure, it has been suggested that interstellar gas including the molecular phase can be pushed up against the centre of a galaxy. This leads to ISM compression and increasing H$_{2}$ formation; hence, it locally triggers intensive star formation \citep{fujita1999,kronberger2008,merluzzi2013,henderson2016}. Indeed, ISM compression with high molecular fraction in the upstream side is seen in a number of galaxies experiencing ICM pressure (\citealt{vollmer2012b}; Nehlig, Vollmer \& Braine 2016), which are often accompanied by an enhancement of star formation. Molecular gas enhancement among the Virgo cluster members has been also recently reported by \cite{mok2016}. These observations clearly show that molecular gas properties and hence star formation activity within a stellar disc can be affected by ram pressure.

Therefore, in order to get a deeper understanding on how galaxies become passive after H{\sc i} stripping, not only the possibility of stripping but also the detailed properties of molecular gas inside a galaxy under strong ICM pressure need to be probed. Hence, using the Submillimeter Array (SMA)\footnote{The Submillimeter Array is a joint project between the Smithsonian Astrophysical Observatory and the Academia Sinica Institute of Astronomy and Astrophysics and is funded by the Smithsonian Institution and the Academia Sinica.}, we have taken high-resolution 12CO (2$-$1) imaging data of a subsample of Virgo galaxies that have lost atomic gas significantly by ICM pressure. Some preliminary results are published in \cite{lee2015}, and in this work, we present more complete analysis on both morphology and kinematics of three Virgo galaxies based on the SMA data.

This paper is organized as follows. We introduce the sample in Section~\ref{sec2}. Details of observations and data reduction procedure are provided in Section~\ref{sec3}. In Section~\ref{sec4}, we present the SMA data, describing the CO morphology and kinematics. In Section~\ref{sec5}, we compare the CO and other wavelength data to discuss how the molecular gas in these galaxies has been affected by the ICM pressure. In Section~\ref{sec6}, we summarize the results and conclude.

A distance of 16 Mpc (1 arcsec $\sim$78 pc) to Virgo cluster is adopted in this work (Yasuda, Fukugita \& Okamura 1997).  

\begin{table*}
 \centering
 \begin{minipage}{100mm} 
  \caption{General Information of Sample Galaxies$^{\rm a}$.}
  \begin{tabular}{@{}lcccc}
  \hline
  Galaxy & NGC~4330 & NGC~4402 & NGC~4522 \\
  \hline
  Right ascension (J2000) 	& $12^{\rmn{h}}23^{\rmn{m}}17^{\rmn{s}}.0$ 
  						  	& $12^{\rmn{h}}26^{\rmn{m}}07^{\rmn{s}}.6$ 
  						  	& $12^{\rmn{h}}33^{\rmn{m}}39^{\rmn{s}}.7$  \\
  Declination (J2000)		& $+11{\rmn{\degr}}22{\rmn{\arcmin}}03{\rmn{\arcsec}}.5$ 
  					  		& $+13{\rmn{\degr}}06{\rmn{\arcmin}}47{\rmn{\arcsec}}.4$ 
  					  		& $+09{\rmn{\degr}}10{\rmn{\arcmin}}30{\rmn{\arcsec}}.2$  \\
  Morphological type & Sc & Sb & SBc \\
  Inclination (${\rmn{\degr}}$) & 79 & 82 & 79 \\
  Position angle (${\rmn{\degr}}$) & 60 & 89 & 35 \\
  $V_{\rmn{rad}} (\rmn{km~s^{-1}})$$^{\rm b}$ & 1565 & 232 & 2328 \\
  $D_{\rmn{25}} (\rmn{arcmin})$ & 2.29 & 3.55 & 3.47 \\
  Total apparent {\it B}-band magnitude & 12.02 & 12.05 & 11.86 \\
  Total {\it K}-band luminosity ($10^{9}{\rm\ L_{\sun,{\it K}}}$)$^{\rm c}$ & 6.58 & 21.30 & 5.64 \\
  $M_{\rmn{HI}}$ ($10^{8} {\rm M_{\sun}}$)$^{\rm d}$ & 4.45 & 3.70 & 3.40 \\
  $def_{\rmn{HI}}$$^{\rm d, e}$ & 0.80 & 0.74 & 0.86 \\
  $d_{\rmn{M87}} ({\rmn{\degr}})$$^{\rm d}$ & 2.1 & 1.4 & 3.3 \\
 \hline
\end{tabular}
 \label{tab1}
$^{a}${General information of the sample galaxies from \cite{paturel2003} (HyperLeda, http://leda.univ-lyon1.fr/).} \\
$^{b}${cf. the Virgo mean $\sim$1100 km~s$^{-1}$ \citep{mei2007}.} \\
$^{c}${\cite{skrutskie2006}, cf. Milky way: $8.24\times10^{10}$ ${\rm L_{\sun,{\it K}}}$ \citep{drimmel2001}, M31: $1.29\times10^{11}$ ${\rm L_{\sun,{\it K}}}$ \citep{barmby2006}.} \\
$^{d}${the VIVA study \citep{chung2009}.} \\
$^{e}${$def_{\rmn{HI}}=\langle \log \overline{\sum}_{\rm HI,all} \rangle - \log \overline{\sum}_{\rm HI,obs}$, where $\langle \log \overline{\sum}_{\rm HI,all} \rangle$ is the mean H{\sc i} surface density of field galaxies \citep{haynes1984}, and $\log \overline{\sum}_{\rm HI,obs}$ is the mean H{\sc i} surface density of an observed galaxy \citep{chung2009}. In this work, morphology independent deficiency has been adopted as \cite{chung2009}.} \\

\end{minipage}

\end{table*}

\section{SAMPLE GALAXIES}
\label{sec2}
The sample for the SMA observations has been selected from the VLA (Very Large Array)\footnote{The National Radio Astronomy Observatory is a facility of the National Science Foundation operated under cooperative agreement by Associated Universities, Inc.} Imaging study of Virgo galaxies in Atomic gas (VIVA) by \cite{chung2009}. The VIVA is a high-resolution H{\sc i} imaging study of 53 late-type galaxies that are located throughout the Virgo cluster from a high-density core region to low-density outskirts. Among the VIVA sample, we have selected three galaxies with clear evidence for active H{\sc i} stripping, NGC~4330, NGC~4402 and NGC~4522 (Fig.~\ref{vir_pos}). Although these galaxies are thought to have lost a similar fraction of H{\sc i} gas, being deficient by a factor of 6$\sim$7 compared to their field counterparts (Table~\ref{tab1}), they show distinct properties in their H{\sc i} morphology.

\begin{figure*}
\centering
\includegraphics[width=1.0\textwidth,bb=15 340 600 500]{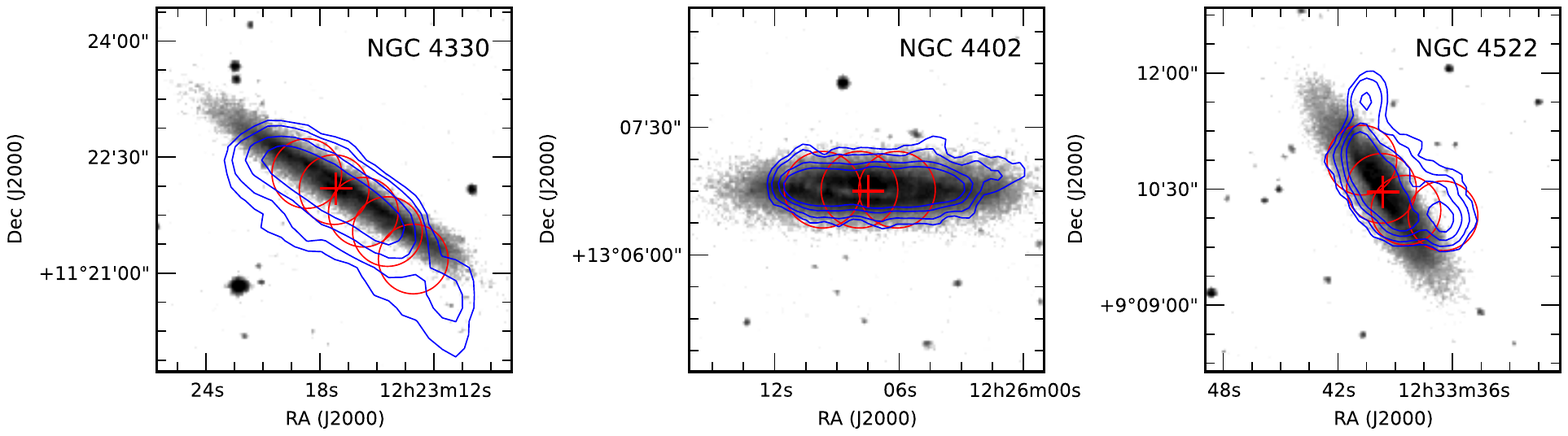}
\caption[]{The H{\sc i} distribution of NGC~4330, NGC~4402 and NGC~4522 (from left to right) is shown in blue contours overlaid on the Digitized Sky Survey 2 (DSS2, https://archive.stsci.edu/dss/index.html) red image. The red cross indicates the stellar disc centre of each galaxy estimated from {\it Spitzer} 3.6~$\mu$m data \citep{salo2015}, and the thin red circles represent the SMA observation points, each of which corresponds to the size of the primary beam at 230~GHz ($\approx$54~arcsec).}
\label{fov}
\end{figure*}

\begin{table*}
\begin{minipage}[h]{120mm}
\caption{Observation parameters.}
\begin{tabular}{@{}lccc}
\hline
Galaxy & NGC~4330 & NGC~4402 & NGC~4522 \\
\hline
Observation date & 2011 Mar 02, 03 & 2010 Mar 20, 21, 27 & 2011 Feb 28, Mar 01 \\
Synthesized beam (arcsec) & 6.35 $\times$ 4.47 & 7.21 $\times$ 3.89 & 6.62 $\times$ 3.92 \\
Position angle (${\rmn{\degr}}$) & -28.4 & 73.4 & -26.4 \\
Spectral resolution ($\rmn{km~s^{-1}}$) & 5.0 & 5.0 & 5.0 \\
Integration time per point (h) & 2.1 & 5.5 & 2.6 \\
rms per channel ($\rmn{mJy\ beam}^{\rmn{-1}}$) \\
~~~~~12CO J$= 2-1$ & 35.2 & 16.3 & 32.7\\
~~~~~13CO J$= 2-1$ $^{\rm a}$ & 29.7 & 13.0 & 29.9 \\
Bandpass calibrators & 0854+201, 1751+096 & 0854+201 & 1751+096, 3c279 \\
Flux calibrators & titan, vesta & mars, titan & mwc349a, vesta \\
Gain calibrators & 3c273, 3c279 & 3c273, 3c279 & 3c273, 3c279 \\
\hline
\end{tabular}
\label{tab2}
{\it Note.} $^{\rm a}${13CO (2$-$1) emission is detected only in NGC~4402.}
\end{minipage}
\end{table*}

NGC~4330 is truncated in H{\sc i} within one side of the stellar disc, while it reveals a long H{\sc i} tail on the opposite side as if the H{\sc i} disc is pushed to the tail side \citep{chung2009, abramson2011}. The location and the H{\sc i} morphology suggest that this galaxy is a recent arrival, entering the high-density region for the first time \citep{chung2007}, and this galaxy will reach the peak pressure after 100 Myr, based on the simulation of Vollmer (2009). Meanwhile, NGC~4402 has been experiencing strong ICM pressure in the last $\sim$150$-$250 Myr \citep{abramson2014}, currently crossing the core region \citep{crowl2005}. Lastly, NGC~4522 is farther away from the cluster centre compared to NGC~4402 in projection, but its H{\sc i} morphology is also suggestive of active ram pressure stripping as NGC~4402. \cite{vollmer2006} show in their simulations that it has been at least 50 Myr, since this galaxy experienced quite strong ICM pressure. This is likely due to turbulence in the ICM that could have been caused by merging of M49 group to the main cluster \citep{kenney2004}.

In spite of subtle differences in their H{\sc i} morphologies, the mean star formation quenching time-scale, i.e. how long ago a galaxy stopped forming stars, is not significantly different among our sample by ranging from 100 to 300 Myr \citep{crowl2008,abramson2011}. Therefore, the detailed CO data of these three cases should enable us to probe how molecular gas properties and thus star formation activities are modified by strong ICM pressure. In addition, by comparing our data with the CO data of NGC~4569, which is known to have crossed the cluster a while ago (Fig.~\ref{vir_pos}, and has a star formation quenching time $\gg$ $\sim$300 Myr; \citealt{vollmer2004, crowl2008}), we will probe how molecular gas properties evolve with time, during the first infall to the cluster and under strong ICM pressure, then after core crossing. For NGC~4569, which is not included in our SMA sample, we make use of the CO data from the HERACLES survey \citep{leroy2009}. The general properties of our SMA sample are summarized in Table~\ref{tab1}.

\section{OBSERVATIONS AND DATA REDUCTION}
\label{sec3}
The SMA is a radio interferometer with eight antenna elements, 6~m each in diameter. It is located in Mauna Kea at the elevation of 4080~m above the sea level. 

Our SMA observations were done in March 2010 and March 2011 in the subcompact configuration. Among eight antennas in total, only seven antennas were available for our observations in both years. The total bandwidth of 4 GHz is composed of 48 chunks that are overlapped by $\sim$20 MHz. Each chunk was configured with 128 channels, each of which is 0.8125~MHz width or 1.1 km~s$^{-1}$ at the rest frequency of 12CO (2$-$1), 230.538~GHz. 12CO for each target was placed on the upper sideband so that 13CO (2$-$1) ($\nu_{\rm rest}=220.398$~GHz) and C18O (2$-$1) ($\nu_{\rm rest}=219.560$~GHz) frequencies were covered simultaneously. The lower sideband where 13CO was included was separated with the upper sideband by 10~GHz.

The primary beam of the SMA is $\sim$54 arcsec at 12CO (2$-$1) rest frequency. Aiming to cover at least half the stellar disc, we mosaicked 3--5 points depending on the optical size of individual galaxies as shown in Fig.~\ref{fov}. In the case of NGC~4522, one additional field in the southwest was included to cover the extraplanar H$\alpha$ and H{\sc i} gas. The total integration time ranges from $\sim$2 to $\sim$5.5~h per field depending on the weather conditions, yet we aimed to achieve a uniform sensitivity for each galaxy. Details of the SMA observations are summarized in Table~\ref{tab2}.

Flux, gain and bandpass calibrations were done using the \textsc{mir} software \citep{qi2012}. The calibrators used for our observations are listed in Table~\ref{tab2}. After the calibration, the data had been analysed using the \textsc{miriad}. The continuum has been subtracted using {\tt UVLIN} by applying a linear fit to the uv-data in line-free channels selected based on the H{\sc i} emission of each galaxy. The imaging and mosaicking were done using {\tt INVERT}. In order to maximize the sensitivity while keeping optimal resolution, the robust was set to 0.5 (for NGC~4402) or 1 (for NGC~4330 and NGC~4522). The channel width in the final cube was regridded to 5 km~s$^{-1}$, which is comparable to H{\sc i} data (VIVA; \citealt{chung2009}). We have detected 12CO (2$-$1) in all three galaxies, while 13CO (2$-$1) has been detected only in NGC~4402, which will be presented in a separate paper (Lee et al. in preparation).

\section{RESULTS}
\label{sec4}
\begin{table*}
\centering
\begin{minipage}{85mm}
\caption{SMA CO properties of sample galaxies.}
\begin{tabular}{lccc}
\hline & NGC~4330 & NGC~4402 & NGC~4522\\ \hline
$W_{\rm 20}$ (km~s$^{-1}$) & 208 & 255 & 176\\
$W_{\rm 50}$ (km~s$^{-1}$) & 184 & 226 & 159\\
$V_{\rm sys}$ (km~s$^{-1}$) & 1562 & 246 & 2326\\
$S_{\rm CO}$ (Jy~km~s$^{-1}$) & 182.22 $\pm$ 8.02 & 1400.76 $\pm$ 11.91 & 139.43 $\pm$ 4.77\\
$M_{\rm{H_{2}}}$ ($10^{8}$ ${\rm M_{\sun}}$)$^{\rm a}$ & 1.19 $\pm$ 0.05 & 8.83 $\pm$ 0.08 & 0.88 $\pm$ 0.03\\
\hline
\end{tabular}
\label{tab3}
{\it Note.} $^{\rm a}$The CO-to-H$_2$ conversion factor of 3.2~${\rm M_{\sun}}$~pc$^{-2}$~(K~km~s$^{-1}$)$^{-1}$ is adopted from \cite{strong1996}.\\
\end{minipage}
\end{table*}


In this section, we present the results of our SMA observations. The 12CO (2$-$1) flux and its uncertainty are measured in Jy~km~s$^{-1}$ as follows,
\begin{equation}
S_{\rm CO} = \sum F_{\rm CO} \times \Delta V \pm
{\left( \sum {{\sigma}^2} \right)}^{\frac {1}{2}} \times \Delta V,
\end{equation}
where $F_{\rm CO}$ is the total flux of CO emission in each channel, $\Delta V$ is the channel separation of the final cube (5 km~s$^{-1}$), and $\sigma$ is the rms of each channel outside CO emission. Continuous features above $2\sigma-3\sigma$ are considered as real signal in individual channels. CO linewidths are calculated using the velocities, where the flux density corresponds to 20 per cent and 50 per cent of the peaks in the receding and the approaching side, adopting the definition of \cite{rhee1996}:
\begin{equation}
W_{\rm 20} = V_{\rm high}^{\rm 20~per~cent} - V_{\rm low}^{\rm 20~per~cent},
\end{equation}
\begin{equation}
W_{\rm 50} = V_{\rm high}^{\rm 50~per~cent} - V_{\rm low}^{\rm 50~per~cent}.
\end{equation}
The CO velocity is determined following \cite{verheijen1997}:
\begin{equation}
V_{\rm sys} = 0.25(V_{\rm low}^{\rm 20~per~cent} + V_{\rm low}^{\rm 50~per~cent} + V_{\rm high}^{\rm 50~per~cent} + V_{\rm high}^{\rm 20~per~cent}).
\end{equation}
 
\begin{figure*}
\begin{center}
\centering
\includegraphics[width=1.0\textwidth,bb=10 120 580 715]{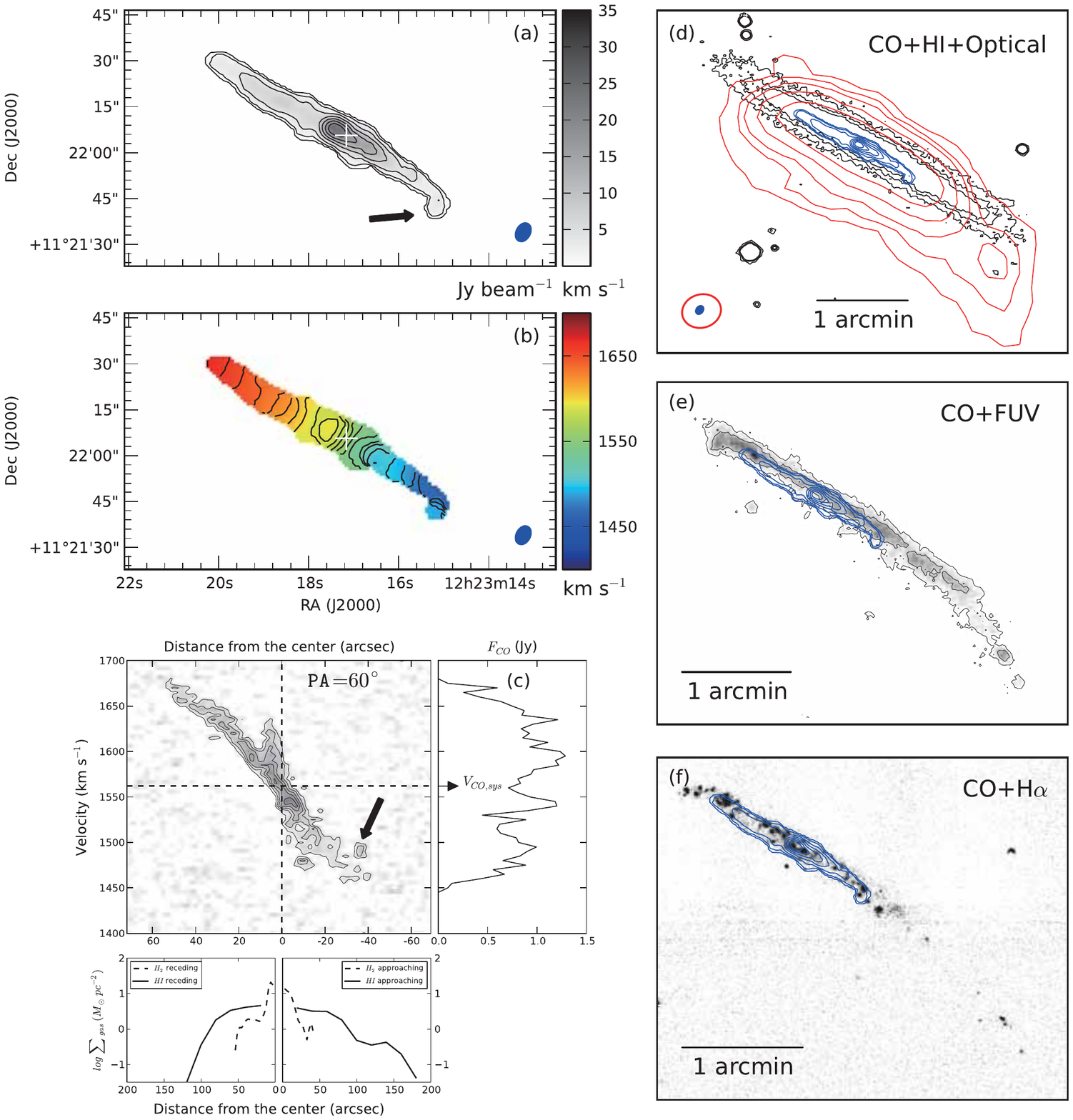}
\caption[]{
NGC~4330: (a) 12CO (2$-$1) integrated intensity map (the 0th moment) in grey-scale with contours. Contour levels are 0.3, 1.5, 4, 8, 12, 16, 20 Jy~beam$^{-1}$~km~s$^{-1}$. The synthesized beam size is 6.35~arcsec $\times$ 4.47~arcsec (blue ellipse at the bottom right). The white cross indicates the stellar disc centre. (b) 12CO (2$-$1) velocity field map (the 1st moment). Velocity contours are drawn in every 10 km~s$^{-1}$ from 1470 km~s$^{-1}$ to 1670 km~s$^{-1}$. The white cross again indicates the stellar disc centre. (c) Upper left: a position-velocity cut through the major axis integrated along the minor axis. Contour levels are 0.7, 1.4, 2.1, 2.8, 3.5 Jy~beam$^{-1}$. The CO clump is indicated by black arrow. Right: the global profile of 12CO (2$-$1). The CO velocity (1562 km~s$^{-1}$) is indicated with an arrow. Bottom: the gas surface densities (H{\sc i} and H$_{2}$) along the approaching side and the receding side on the right and the left, respectively. (d) An overlay of 12CO (2$-$1) (blue contours) on H{\sc i} (red contours) and DSS2 red (black contours). Synthesized beam of the VLA and the SMA are shown in red and blue at the bottom left. The entire CO disc is located inside the optical disc. Unlike the stellar disc, however, the CO disc is found to be highly asymmetric as the atomic gas disc. (e) 12CO (2$-$1) (blue contours) is overlaid on FUV emission (black contours and grey-scale). FUV tail is also extended and bent toward the south-west as CO and H{\sc i} gas tail. (f) 12CO (2$-$1) (blue contours) overlaid on the H$\alpha$ emission (grey-scale). The overall bending shape coincides well between H$\alpha$ and CO, while CO is not extended as much as H$\alpha$.
}
\label{n4330}
\end{center}
\end{figure*}

\begin{figure*}
\begin{center}
\includegraphics[width=1.0\textwidth,bb=10 120 580 715]{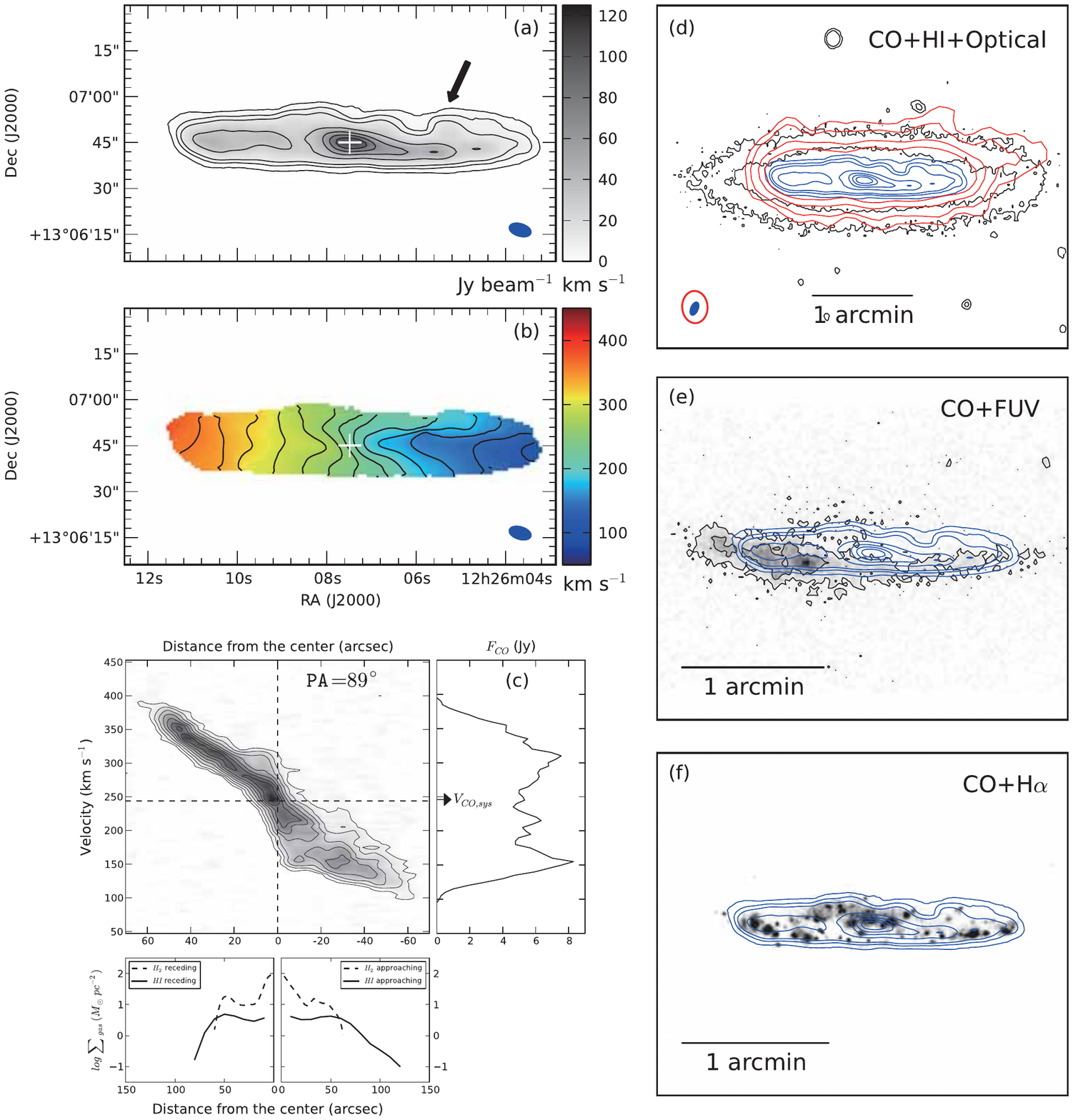}
\caption[]{
NGC~4402: (a) 12CO (2$-$1) integrated intensity map (the 0th moment) in grey-scale with contours. Contour levels are 1, 5, 10, 30, 50, 70, 90 Jy~beam$^{-1}$~km~s$^{-1}$. The synthesized beam size is 7.21~arcsec $\times$ 3.89~arcsec (blue ellipse at the bottom right). The white cross indicates the stellar disc centre. A modest CO bump is indicated by black arrow. (b) 12CO (2$-$1) velocity field map (the 1st moment). Velocity contours are drawn in every 20 km~s$^{-1}$ from 120 km~s$^{-1}$ to 360 km~s$^{-1}$. The white cross again indicates the stellar disc centre. (c) Upper left: a position-velocity cut through the major axis integrated along the minor axis. Contour levels are 0.5, 1.5, 3, 5, 7, 9, 11 Jy~beam$^{-1}$. Right: the global profile of 12CO (2$-$1). The CO velocity (246 km~s$^{-1}$) is indicated with an arrow. Bottom: the gas surface densities (H{\sc i} and H$_{2}$) along the approaching side and the receding side on the right and the left, respectively. (d) An overlay of 12CO (2$-$1) (blue contours) on H{\sc i} (red contours) and DSS2 red (black contours). 12CO gas is well confined within the optical disc, while HI is pushed off outside from the stellar disc. The optical disc does not look disturbed, while both molecular and atomic gas components reveal asymmetry in a similar sense. (e) 12CO (2$-$1) (blue contours) is overlaid on FUV emission (black contours and grey-scale). FUV is found to be enhanced along the CO compression. (f) 12CO (2$-$1) (blue contours) overlaid on the H$\alpha$ emission (grey-scale). The overall extent and morphology well coincide with each other.
}
\label{n4402}
\end{center}
\end{figure*}

\begin{figure*}
\begin{center}
\includegraphics[width=1.0\textwidth,bb=10 120 580 715]{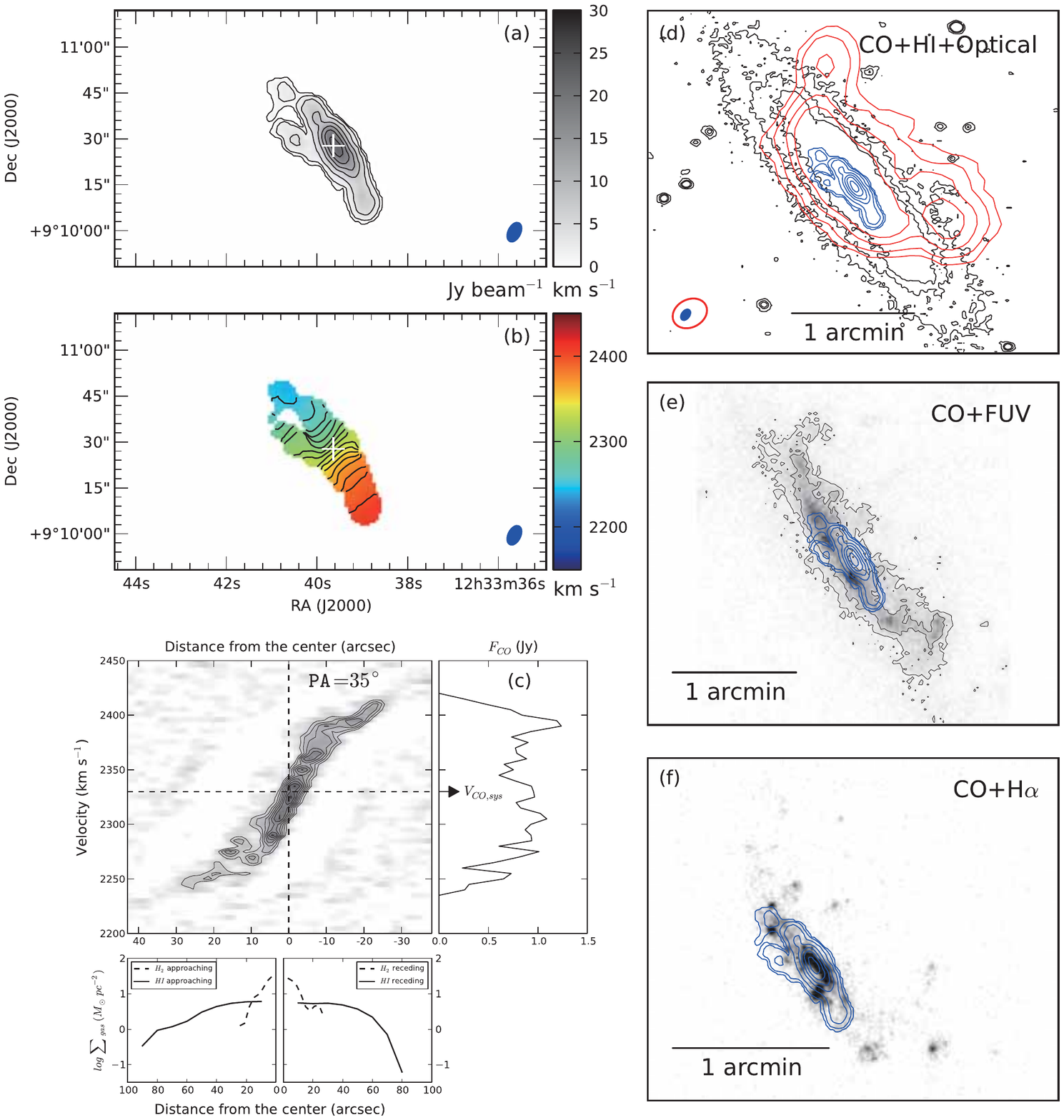}
\caption[]{
NGC~4522: (a) 12CO (2$-$1) integrated intensity map (the 0th moment) in grey-scale with contours. Contour levels are 0.3, 1.5, 4, 8, 12, 16, 20 Jy~beam$^{-1}$~km~s$^{-1}$. The synthesized beam size is 6.62~arcsec $\times$ 3.92~arcsec (blue ellipse at the bottom right). The white cross indicates the stellar disc centre. (b) 12CO (2$-$1) velocity field map (the 1st moment). Velocity contours are drawn in every 10 km~s$^{-1}$ from 2250 km~s$^{-1}$ to 2400 km~s$^{-1}$. The white cross again indicates the stellar disc centre. (c) Upper left: a position-velocity cut through the major axis integrated along the minor axis. Contour levels are 1.5, 2.0, 2.5, 3.0, 3.5, 4.0, 4.5, 5.0 Jy~beam$^{-1}$. Right: the global profile of 12CO (2$-$1). The CO velocity (2326 km~s$^{-1}$) is indicated with an arrow. Bottom: the gas surface densities (H{\sc i} and H$_{2}$) along the approaching side and the receding side on the left and the right, respectively. (d) An overlay of 12CO (2$-$1) (blue contours) on H{\sc i} (red contours) and DSS2 red (black contours). Our SMA data reveal only the inner part of the molecular gas disc, missing out the outer part that is detected by the IRAM \citep{vollmer2008} due to the lack in pointings and sensitivity. In Section~\ref{sec4} and~\ref{sec5}, we more focus on the morphology and kinematics revealed by the SMA. (e) 12CO (2$-$1) (blue contours) is overlaid on FUV emission (black contours and grey-scale). Enhanced FUV emission is found along the side on which the ICM wind is presumably acting. (f) 12CO (2$-$1) (blue contours) overlaid on the H$\alpha$ emission (grey-scale). The inner CO disc is slightly bent toward south-east as the inner H$\alpha$ disc.
}
\label{n4522}
\end{center}
\end{figure*}

Following the recipe from \cite{solomon1997}, 12CO (2$-$1) line luminosity is measured as follows:
\begin{equation}
L_{\rm CO}^{'}(2-1)=3.25\times10^7 S_{\rm CO}~\nu_{\rm obs}^{-2}~D_{L}^2~(1+\textit{z})^{-3}, 
\end{equation}
in $\rm{K~km~s^{-1}~pc^2}$, where $S_{\rm CO}$ is an integrated total 12CO (2$-$1) flux in Jy~km~s$^{-1}$, $D_{L}$ is the luminosity distance in Mpc, $\nu_{\rm obs}$ is the observation frequency in GHz, and \textit{z} is the redshift. Adopting a typical 12CO (2$-$1)/(1$-$0) ratio of nearby galaxies ($R_{\rm 21}$$\approx$0.8; \citealt{leroy2009}) and using the conversion relation of normal spirals ($\alpha_{\rm CO}$~=~3.2~${\rm M_{\sun}}$~pc$^{-2}$~(K~km~s$^{-1}$)$^{-1}$;  \citealt{strong1996}), the molecular gas mass can be calculated by 
\begin{equation}
M_{\rm H_{2}} = {\frac {\alpha_{\rm CO}}{R_{\rm 21}}}~L_{\rm CO}^{'}(2-1),
\end{equation}
where $L_{\rm CO}^{'}$(2$-$1) is the luminosity of 12CO (2$-$1) transition.
The global CO properties of the sample are summarized in Table~\ref{tab3}. On the left of Figs~\ref{n4330}--\ref{n4522}, the CO intensity map, velocity field and position-velocity diagram (PVD) with global and radial profile are presented in (a), (b) and (c) for each galaxy. The overlays with other wavelength data shown on the right-hand side of Figs~\ref{n4330}--\ref{n4522} are discussed in Section~\ref{sec5}.


\subsection{NGC~4330}
As seen in Figs~\ref{n4330}(a)--(c), the CO morphology of NGC~4330 is highly asymmetric, with the south-west extent in the downstream side being about 75 per cent (40 arcsec versus 53 arcsec) of the northeast but only 68 per cent when the outermost 4 arcsec of the south-west disc, which is bent down to the south (indicated by arrow in Fig.~\ref{n4330}a) is excluded (36 arcsec versus 53 arcsec). The central part, within 30 arcsec in diameter ($\sim$2 kpc at the distance of Virgo), is also asymmetric due to the strong local peak in the north-east from the stellar disc centre. The distinct extents and surface densities are also clearly seen in radial profiles. In the end of the southwest disc of the downstream side, CO is found to be slightly bent (Fig.~\ref{n4330}a). This bending is also found in many other wavelengths such as UV, H$\alpha$ and H{\sc i}, but all are different in scale and angle from one another as further discussed in Section~\ref{sec5}. In the case of CO, this bending part is clumpy, almost identified as an independent blob or clump (see the arrow in Fig.~\ref{n4330}c). This CO clump corresponds to 2 per cent of the total in flux ($\sim$1.9$\times$10$^6$~${\rm M_{\sun}}$, comparable to that of a large molecular cloud). The centre of the clump is off from the mid-plane of the main disc toward the south by $\sim$4 arcsec ($\sim$312~pc).

The CO kinematics also shows peculiar structures, especially along the south-west concentration and the end of the tail as shown in Fig.~\ref{n4330}(b).  The velocity gradient of the south-west is quite steep, while it is more slowly rising on the other side within the small radii from the centre, reaching 72 km~s$^{-1}$ on the approaching side but only 43 km~s$^{-1}$ on the receding at 15 arcsec radii. In the south-west clump, the velocity gradient is inverted. The kinematical complexity is also clearly seen in the PVD (Fig.~\ref{n4330}c). A distinct component of a steep velocity gradient in the central $\sim$13 arcsec ($\sim$1 kpc) is quite noticeable. This may indicate a molecular ring or bar. In addition, the CO peak is off from the stellar disc centre by $\sim$270 pc (Fig.~\ref{n4330}a), while the CO kinematic centre is more or less consistent with the stellar disc centre (Fig.~\ref{n4330}c).

\subsection{NGC~4402}
As shown in Fig.~\ref{n4402}(a), the CO disc of NGC~4402 is slightly more extended in the west (62 arcsec in the west versus 58 arcsec in the east), but the difference is subtle and not as significant as in H{\sc i}. What makes this case look highly asymmetric is the north-west quarter of the downstream side (the west side of the disc). In this region, a modest CO bump (indicated by arrow in Fig.~\ref{n4402}a) is found as in H{\sc i} (Fig.~\ref{n4402}d). Meanwhile, the southern part of the CO disc looks quite compressed along the enhanced far-ultraviolet (FUV) emission (Fig.~\ref{n4402}e), which will be discussed more in detail in Section~\ref{sec5}. The northern part above the major axis is measured to be 91 per cent of that of the south in flux. Although the CO peak agrees well with the stellar disc centre, the inner CO disc also appears to be slightly more stretched toward the west, reflecting the outer CO disc.

The CO kinematics reveals the evidence for non-circular motions as shown in Fig.~\ref{n4402}(b). In the inner region, the iso-velocity curves are highly skewed and not perfectly parallel with the minor axis. The skewness in velocity must be the result of two distinct disc components in the inner $\sim$15 arcsec ($\sim$1 kpc) being projected on the sky as seen in its PVD (Fig.~\ref{n4402}c). The velocity gradient in the outer part is quite different in the two sides of the disc along the major axis. In the north-west quarter (in the downstream side), the kinematics of modest CO bump is slightly deviated from the main disc.

It also has a distinct CO component with steep velocity gradient in the inner $\sim$15 arcsec. This likely indicates a nuclear bar or ring, although we favor a nuclear bar as this would also explain the skewed isovelocity contours. In NGC~4402, \cite{sofue2003a, sofue2003b} also find a nuclear molecular disc traced by 12CO (1$-$0) in the inner $\sim$10 arcsec. We do not find any direct relevance of these nuclear structures with ram pressure stripping.

\subsection{NGC~4522}
While the overall size of 12CO (2$-$1) disc in the SMA image is consistent with the extent measured by a single dish for NGC~4330 \citep{vollmer2012a} and NGC~4402 (Lee et al. in preparation), our SMA data of NGC~4522 reveal only the inner $\sim$57 per cent of its single-dish map in size \citep{vollmer2008}. While \cite{vollmer2008} have detected CO along the extraplanar H$\alpha$ and H{\sc i} in both ends of the disc, we did not detect any such features in the south-west in our SMA data due to the lack of sensitivity and/or potentially due to the diffuse nature of gas in the outer molecular disc. The north-east end was not covered in the SMA observations. Therefore, we limit our discussion to only the inner CO disc for this particular case in this section and we cite the IRAM single-dish data when it is needed for comparisons with other wavelength data in Section~\ref{sec5}.

\begin{figure*}
\begin{center}
\includegraphics[width=1.0\textwidth,bb=15 165 580 680]{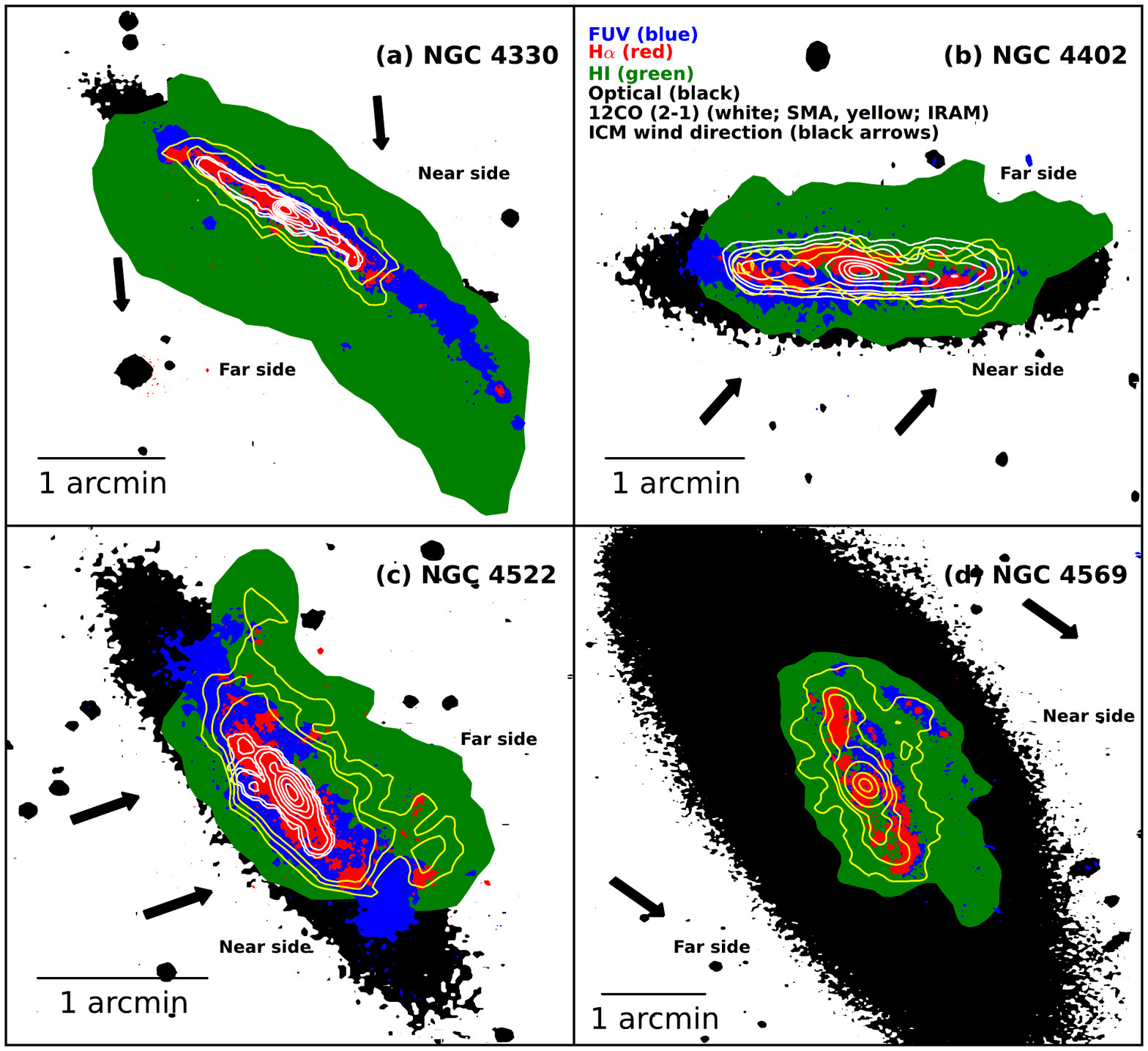}
\caption[]{A composite map of FUV (blue; \citealt{gil2007}), H$\alpha$ (NGC~4330 and NGC~4569 from \citealt{gavazzi2003}, NGC~4402 from \citealt{crowl2005}, NGC~4522 from \citealt{koopmann2001}), optical (black; DSS2 red), H{\sc i} (green; \citealt{chung2009}) and 12CO (2$-$1) (white contours from our SMA data) of NGC~4330 (redshifted), NGC~4402 (blueshifted), NGC~4522 (redshifted), and NGC~4569 (blueshifted). The IRAM 12CO (2$-$1) data are shown in yellow contours. Contour levels are 3.7, 7.5 Jy~beam$^{-1}$~km~s$^{-1}$ for NGC~4330 \citep{vollmer2012a}; 5, 20 Jy~beam$^{-1}$~km~s$^{-1}$ for NGC~4402 (Lee et al. in preparation); 1.7, 3.5, 5.3 Jy~beam$^{-1}$~km~s$^{-1}$ for NGC~4522 \citep{vollmer2008} and 14, 71, 106, 214, 498, 712 Jy~beam$^{-1}$~km~s$^{-1}$ for NGC~4569 \citep{leroy2009}. The ICM wind direction deduced from H{\sc i} morphology is shown with black arrows \citep{vollmer2004, crowl2005, abramson2011, abramson2014}. NGC~4330, NGC~4402 and NGC~4522 are at relatively early to active H{\sc i} stripping stage due to the ICM, while NGC~4569 is thought to be a post-peak pressure case where some H{\sc i} gas is falling back on the disc after core crossing.}
\label{multi}
\end{center}
\end{figure*}

As shown in Fig.~\ref{n4522}, the CO extent is comparable in both sides (27.4 arcsec in the north-east versus 27.6 arcsec in the south-west). The inner CO disc of NGC~4522 is found to be curved in the opposite way to the outer CO and H{\sc i} disc, i.e. to the south-east as the inner H$\alpha$ disc. We find two features sticking out from the inner part, one from the end of north-east disc and another almost to the same direction but connected from the centre of the main disc, more to the south. The north-east blob coincides with the morphology of the inner CO disc in the IRAM image (Fig.~\ref{multi}c; \citealt{vollmer2008}), which is smoothly connected to the disc farther extended in the north-east, coinciding with a dust loop in this region. While the downward bending coincides with the inner spiral structure including the dust feature (see Fig.~\ref{hst} in Appendix A). 

The CO velocity structure within the main disc generally shows that of a regularly rotating disc as seen in Figs~\ref{n4522}(b) and (c). The velocity keeps rising up to $\sim$10 arcsec on both sides, then almost flattens out, and toward the end of the disc, the velocity rises again. The velocity gradient along the north-east branch is overall consistent with that of the main disc on the same side. However, the gradient is smaller and the velocity rises more slowly compared to the other side.

\begin{figure*}
\begin{center}
\includegraphics[width=1.0\textwidth,bb=5 245 600 590]{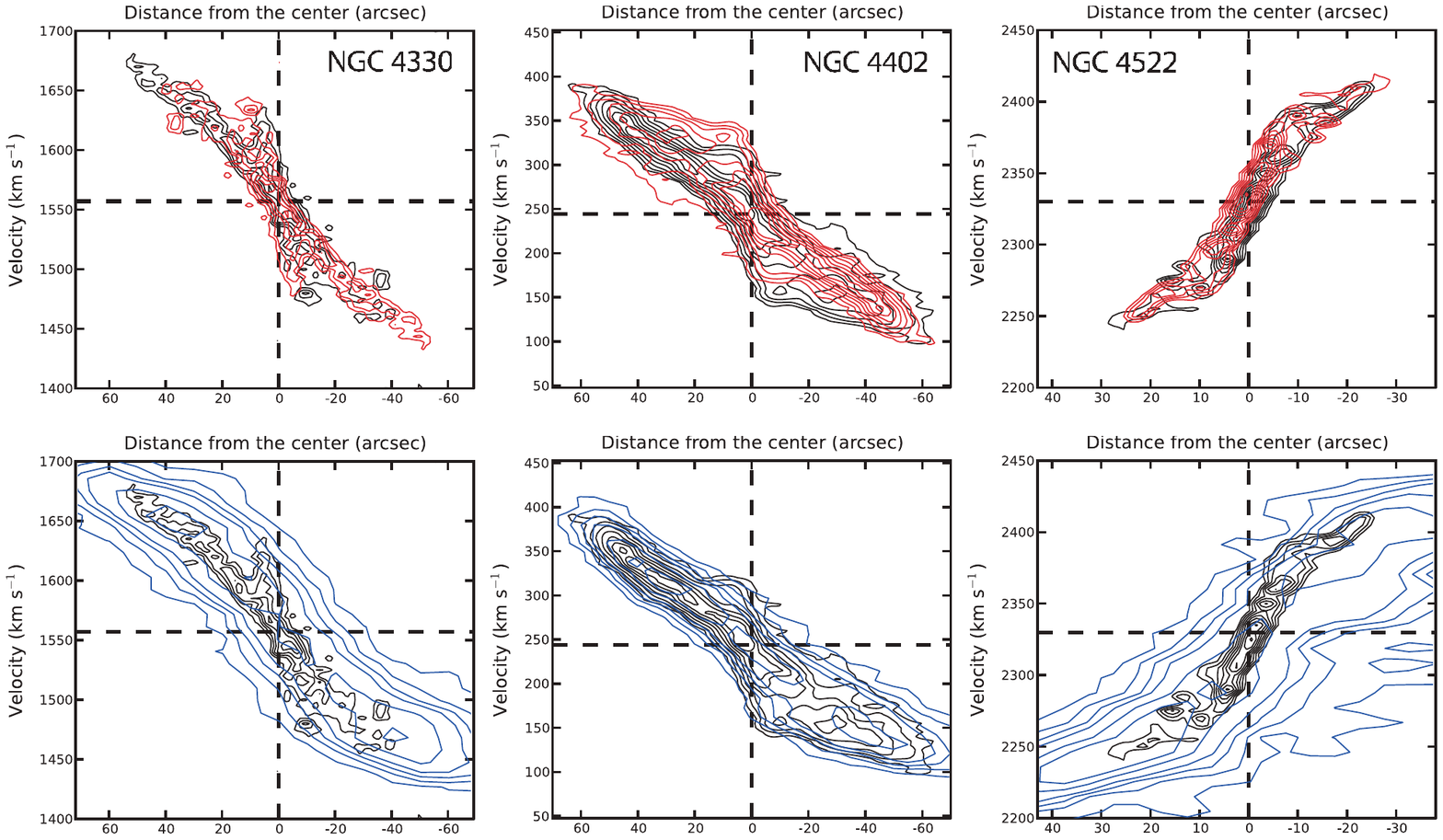}
\caption[]{The PVD of each galaxy (black contours) is mirrored (red contours) and overlaid on the original one. These images are folded along the centre of the stellar disc. The kinematics of molecular gas in NGC~4330 and NGC~4402 is quite distinct in the receding and the approaching side. In the bottom row, the comparisons with the H{\sc i} PVD (blue contours) are presented.}
\label{pvd}
\end{center}
\end{figure*}

\section{DISCUSSION}
\label{sec5}
\subsection{Comparison with other wavelength data}
In this section, we compare 12CO (2$-$1) properties with other wavelength data including H{\sc i} \citep{chung2009}, optical (DSS2 red), FUV \citep{gil2007}, and H$\alpha$ (NGC~4330 and NGC~4569 from \citealt{gavazzi2003,abramson2011}, NGC~4402 from \citealt{crowl2005}, NGC~4522 from Koopmann, Kenney, \& Young 2001). CO traces star-forming gas, and H$\alpha$ and FUV emission are good indicators of star formation with time-scales of $\sim$20 and 100 Myr, respectively \citep{kennicutt1998}. We also compare the radial gas surface density between atomic and molecular hydrogen. The H{\sc i} column density is calculated using,
\begin{equation}
N({\rm HI}) = 1.82 \times 10^{18} \times \sum T_{\rm B}\triangle V
\end{equation}
in cm$^{-2}$, where $\sum T_{\rm B}\triangle V$ is the integrated intensity (brightness temperature) in $\rm K~km~s^{-1}$ \citep{walter2008}. The H$_{2}$ surface density is estimated by the following relation:
\begin{equation}
N({\rm H_{2}}) = {\frac {\alpha_{\rm CO}}{R_{\rm 21}}}~I({\rm CO})
\end{equation}
in ${\rm M_{\sun}}$~pc$^{-2}$, where $I(\rm CO)$ is 12CO (2$-$1) integrated intensity, and ${\frac {\alpha_{\rm CO}}{R_{\rm 21}}}$ is the same as in equation (6). Considering the high inclination of all three galaxies in our sample, we utilize the strip integral method instead of ellipse fitting to derive radial surface density because ellipse fitting is usually inappropriate for edge-on galaxies when the spatial resolution is limited \citep{warmels1988a, rhee1996, swaters1999}. This method calculates {\em strip integrals} by integrating H{\sc i} column density perpendicularly to the major axis, using Lucy's (1974) iterative deconvolution method. Then the strip integrals of approaching and receding sides are deprojected separately to infer the face-on H{\sc i} surface density \citep{kregel2004}. The surface density of both atomic and molecular hydrogen is measured in ${\rm M_{\sun}}$~pc$^{-2}$. The comparison between atomic and molecular gas radial distributions is found in the bottom on the left-hand column of Figs~\ref{n4330}--\ref{n4522}. Overlays between CO and other multiwavelength data are presented on the right-hand column of Figs~\ref{n4330}--\ref{n4522} and Fig.~\ref{multi}.

For the reference, the IRAM 12CO (2$-$1) maps are also overlaid in the composite map. The sensitivity of the IRAM cubes is $\sim$10 mK per 5 km~s$^{-1}$ channel (the same channel width as ours), which is comparable to the SMA cubes or slightly better in the cases of NGC~4330 and NGC~4522 when the different spatial resolutions are taken into account ($\sim$11 arcsec versus $\sim$5 arcsec for IRAM and SMA, respectively). For NGC~4402, the sensitivity of SMA data is somewhat better than the IRAM data not only because of the sufficient integration time with the SMA but also because of the poor weather condition when the IRAM data were taken. For NGC~4330 and NGC~4402, the IRAM and SMA maps are generally in good agreement in morphology, considering the beam sizes, while we are missing most of the extraplanar molecular of NGC~4522 in the SMA data due to the lack in coverage, sensitivity, and/or potentially due to the diffuse nature of molecular gas in the extragalactic space as further discussed in Section 5.1.3.

\subsubsection{NGC~4330}
Asymmetry is ubiquitous in a range of wavelengths data except in the red optical image (DSS2 red). In Fig.~\ref{multi}(a), we see that the H{\sc i} is truncated within the stellar disc in the north-east, where H$\alpha$ and FUV emission reveals an upturn feature \citep{chung2009, abramson2011}. A hint of upturn is also found in the north-east end of the upstream of radio continuum at 6 cm and 20 cm \citep{chung2009, vollmer2012a}. The single-dish 12CO (2$-$1) data \citep{vollmer2012a} appear to be similar to the radio morphology, also with some hint of upturn, although the extent of the molecular gas is much less than that of H{\sc i}. In our SMA data, we are missing some outermost features including the tip of the nose in the north-east. Instead, the SMA data clearly reveal very detailed structures of the inner CO disc such as the difference in scaleheight between the north and the south along the major axis (up to $\sim$0.36 kpc, Fig.~\ref{n4330}a) which is not as clear as this in the previous single-dish data.

The south-west side, i.e. the downstream of the galaxy where an H{\sc i} gas tail is present, is bent toward the south. On this side, we also find FUV, H$\alpha$ and radio continuum tail but the extents and bending angles are all different in various wavelengths (Fig~\ref{multi}a). The distribution becomes patchy in the outer CO disc as seen in both IRAM single-dish data \citep{vollmer2012a} as well as our SMA data. Intriguingly, the CO blob toward the end is located in almost the same galactic radius as one of the distinct H$\alpha$ blobs and off from the mid-plane in the same direction as the tails of the other wavelengths (Fig~\ref{multi}a).

As clearly seen, the molecular disc, which extends to only half the stellar disc or less, shows many properties in common with the morphologies in the other wavelengths, reflecting the impact of the ICM pressure. Although there are subtle differences, this suggests that the molecular gas in this galaxy has been also affected in similar ways by the same mechanism that is responsible for the peculiarities found in the other wavelengths. This indicates that ICM pressure can change molecular gas properties well inside the stellar disc.

\subsubsection{NGC~4402}
The CO emission shows a number of properties in common with the H{\sc i} gas (Fig.~\ref{n4402}d). It is somewhat extended in the west, i.e. the same side where the H{\sc i} is more extended. Both phases show a bump in the west (indicated by arrow in Fig.~\ref{n4402}a), although the scale is distinct because the H{\sc i} bump is visible in extraplanar gas. H{\sc i} is truncated within the mid-plane of the stellar disc on both sides, with a sharp cut-off in the east to south along the upstream. Meanwhile, the northern side is more extended with a short tail pointing the north-west in the downstream side. The extent of the molecular gas is not significantly different in the receding and the approaching side, yet its distribution is quite distinct. These differences between the two sides of the CO disc can be also clearly seen in the radial velocity distributions (Fig.~\ref{n4402}c). Note that the H{\sc i} extent appears to be almost twice as long as the CO in the west due to the tail (Figs~\ref{n4402}c and d).

The molecular gas disc is thicker in the east (especially third contour level from the outermost contour), reflecting the change in FUV disc thickness across the disc. Although the overall morphology of FUV is quite different from that of CO, both the FUV enhancement and CO compression are present in the east and south-east (Fig.~\ref{n4402}e). H$\alpha$ is tightly correlated with CO in general, with a similar distribution and extent. Also, strong H$\alpha$ knots are found near or within the local CO peaks (Fig~\ref{n4402}f). As with NGC~4330, the CO morphology is surprisingly similar to that of H{\sc i}, strongly suggesting that the impact of the ICM has reached the inner interstellar gas disc.

\subsubsection{NGC~4522}
With given sensitivity limit, we are missing the outer part of the molecular gas disc in the SMA data for this case. As shown by \cite{vollmer2008}, however, the overall morphology of CO is quite similar to that of H{\sc i}, revealing a significant fraction of molecular gas outside the stellar disc (Fig~\ref{multi}c). This galaxy must be one of the best examples found to date where extraplanar molecular gas is present. In addition to the sensitivity, the molecular gas could be more diffuse in nature compared to the gas within the disc, which could make the detection more difficult. In fact, we recently have detected some extragalactic 13CO (1$-$0) in NGC~4522 using the Atacama Large Millimeter Array (ALMA) - cycle 3. The nature of this extraplanar CO gas will be further analyzed using 12/13CO ratio, and will be presented in a separate work.

The inner CO disc is slightly bent down to the south as H$\alpha$ emission. The strongest H$\alpha$ blobs near the galactic centre are located within the inner CO disc (Fig~\ref{n4522}f), but there is an offset between the inner FUV enhancement and the inner CO disc (Fig~\ref{n4522}e). However, this offset in the inner region might be simply due to the dust extinction effect. On the other hand, the offset between CO (the IRAM data)/H$\alpha$ and FUV at large radii is real since there is almost no dust beyond the CO extent. The extent of FUV is larger than the CO and H$\alpha$ extents within the stellar disc by a factor of 1.5 (Fig.~\ref{multi}c). However, the FUV enhancement is roughly consistent with the CO compression in the downstream side as in NGC~4402. This implies that there is significant change in the recent star formation activity and the distribution of molecular gas in a very short time as further discussed in Section~\ref{sec5}.3.

\subsection{Impact of ram pressure on molecular gas}

Most previous studies using single-dish radio telescopes did not find a significant difference in CO luminosities between field galaxies and cluster members \citep{stark1986, kenney1989}. However, some individual galaxies in the cluster environment have been reported to show very distinct CO distributions from those with normal H{\sc i} content in the field (e.g. \citealt{kenney1990, sofue2003c}). More recently, \cite{fumagalli2009} have shown that the molecular gas surface density of galaxies that are H{\sc i}-deficient tends to be low. These suggest that the molecular gas of cluster galaxies can be affected in some ways by ram pressure even if it is not stripped. Indeed, our high-resolution CO data clearly show that molecular gas has been affected in a similar manner as diffuse gas by ram pressure, and the impact of ICM winds reaches as deeply as a few kpc from the galactic centre, as we further elaborate below.

Firstly, the CO morphology in our sample is all found to be highly asymmetric and disturbed. The inner molecular gas disc of many field galaxies does show some peculiarities to a certain degree, including asymmetry (e.g. BIMA SONG survey; \citealt{helfer2003}). However, the overall asymmetry found in the CO morphology of all three galaxies is similar to that of the H{\sc i} morphology. In Figs~\ref{multi}(a)--(c), the composite maps of cool ISM contents (CO and H{\sc i}), star formation indicators (FUV and H$\alpha$) and the old stellar disc (DSS2 red) of our SMA sample are presented. A rough direction of the ICM wind based on the H{\sc i} morphology for each case is shown with thick arrows. 

Secondly, the detailed CO structure of the sample shows good connections with the H{\sc i} features which have been pushed and/or stripped by the ICM pressure. In NGC~4330, the CO clump in the south-west end seems to have been pushed down to the south from the mid-plane of the stellar disc, similarly to its H{\sc i} tail (Fig.~\ref{n4330}d). In addition, similar kinematical structures are found in this side, i.e. in the downstream in both H{\sc i} and CO (Fig.~\ref{pvd}). For example, the CO gas reveals an inverted velocity gradient in the end of south-west like H{\sc i} which shows an anomalous velocity component in the same side \citep{abramson2011}. This implies that dense molecular phase gas can be pushed by ram pressure in addition to the diffuse atomic phase gas. Although the morphology is suggestive of gas being pushed away, one problem with this possibility is that both the H{\sc i} tail and the CO clump in the downstream are found with higher velocities than the main disc, i.e. the opposite direction of the system velocity of cluster centre. This is inconsistent with what is generally expected in the case of the gas pushed by the ICM as found in other Virgo spirals (e.g. NGC~4522; Kenney, van Gorkom \& Vollmer 2004). However, this is still feasible if the gas has been losing angular momentum in extraplanar space as suggested by \cite{abramson2011}.

Alternatively, the CO clump might be an example of a dense cloud unveiled after the stripping of surrounding diffuse gas, which hence appears to stand alone, decoupled from the rest of the disc, similar to the ones seen as narrow dusty plumes found in Virgo galaxies such as NGC~4402 and NGC~4522 \citep{crowl2005,abramson2014,kenney2015} or NGC~4921, a Coma spiral \citep{kenney2015}. If this is an example of those decoupled clouds, it is the first time it has been seen clearly in CO. However, one aspect that makes this case particularly interesting is its gas kinematics. Rather than being smoothly connected from the inner disc, the velocity gradient at the location of the CO clump turns back. This velocity structure strongly suggests that this clump is the molecular gas that has been pushed off from the main disc.

In NGC~4402, both H{\sc i} and CO are compressed from the south-east to south along the upstream. Meanwhile in the north-west, a modest CO bump is found to be pointing to the same direction as the H{\sc i} bump (Fig.~\ref{n4402}d). The compressed CO in the upstream side reveals a steeper velocity gradient than the other side. Meanwhile, the CO bump in the north-west downstream side (Fig.~\ref{n4402}b) is found to have larger velocities than the main disc (Fig.~\ref{n4402}c), i.e. closer to the cluster centre. This is consistent with our expectations if the gas is accelerated by ram pressure due to the ICM.

In NGC~4522, the overall CO morphology from the IRAM data \citep{vollmer2008} is very similar to the H{\sc i} morphology, with the extraplanar gas in the downstream side and compression in the upstream side (Fig.~\ref{multi}c). This strongly suggests that the molecular gas has been influenced by the same mechanism that affects the atomic gas, i.e. ram pressure due to the ICM wind. In the upstream side, the interstellar gas can also get compressed by the ICM pressure, which potentially results in the formation of molecules as inferred by the excess of CO in this side (see also the example of NGC~4654;  \citealt{chung2014}).

As shown in Fig.~\ref{pvd}, the CO kinematics also suggests that the ICM pressure can disturb the molecular gas, deep inside a galactic potential well. In NGC~4330 and NGC~4402, the CO velocity gradients are quite distinct in the receding and the approaching side, an asymmetry also observed in H{\sc i}. In the outer part of NGC~4330 and NGC~4402 (the CO clump and the modest CO bump, respectively) where the molecular gas is off from the mid-plane, we also find the velocity gradient deviates from the overall gas flow of the gas disc. The comparisons with the H{\sc i} PVD is also intriguing (the bottom row of Fig.~\ref{pvd}). In the upstream of NGC~4330 and NGC~4402, the velocity gradients of the molecular gas and atomic gas are similar to each other, but they deviate in the downstream. In NGC~4522, the deviation starts at $\sim$0.8 kpc from the centre on both sides. This may indicate a delay in stripping of gas in distinct phases, i.e. more diffuse gas dragged prior to relatively dense gas.

Our results clearly show that the molecular gas in galaxies experiencing active H{\sc i} stripping can become highly asymmetric, as demonstrated by simulations (e.g. \citealt{hidaka2002, vollmer2008, vollmer2012a}). In addition, our high-resolution data indicate that the influence of strong ICM pressure reaches quite deep inside of the galactic disc. Although molecular gas may be pushed off from the stellar disc together with atomic gas as in NGC~4522 \citep{vollmer2008}, active ram pressure does not seem to always cause molecular gas stripping (NGC~4330 and NGC~4402). Indeed, NGC~4569, which is thought to have lost more than 90 per cent of its H{\sc i} gas while crossing the cluster core a few hundred Myr ago (H{\sc i} deficiency: 1.47, a projected distance of 1.7 degree; $\sim$0.5 Mpc to M87, the bottom right of Fig.~\ref{multi}), has a comparable molecular gas fraction to optical luminosity of our SMA sample (NGC~4330, NGC~4402 and NGC~4522). Adopting the H$_{2}$ deficiency of \cite{fumagalli2009}, single-dish measured CO fluxes and the MW conversion factor result in $M({\rm H_{2}})_{\rm observed} / M({\rm H_{2}})_{\rm expected}$ of 0.48, 1.25, 0.97 and 1.19 for NGC~4330, NGC~4402, NGC~4522 and NGC~4569, respectively. Considering that the sample of \cite{fumagalli2009} does not include galaxies as faint as NGC~4330 or NGC~4522, the expected H$_{2}$ mass could have been overestimated for these two galaxies. This supports that molecular gas mass does not necessarily change significantly during and after severe H{\sc i} stripping, while its detailed molecular gas properties may change. These changes to the CO gas properties are consistent with those from \cite{wong2014}, who found enhanced levels of warm molecular gas which resulted from additional ISM-heating due to ram pressure driven shocks in NGC~4330, NGC~4402 and NGC~4522.

\subsection{Evolution of molecular gas and star formation activities by ram pressure}
\subsubsection{Galactic disc}

All three galaxies have lost H{\sc i} gas significantly, containing only $<$~$\sim$20 per cent of H{\sc i} mass compared to field galaxies of similar size. The star formation quenching time-scale, $\tau_{\rm q}$, measured based on the youngest stellar age at the location, where H{\sc i} gas is truncated (i.e. one way to estimate how long ago star formation was suppressed in a galaxy), ranges from a few tens Myr to a few hundreds Myr for our sample \citep{crowl2008,abramson2011}. It is however comparable among the sample within the uncertainties, and they share several properties in common in star formation activity in spite of subtle differences due to the ICM wind angle and the projection on the sky as shown in Fig.~\ref{multi}.

The spatial distributions of CO and H$\alpha$ (a tracer of recent star formation with a timescale of $<\sim$20 Myr; \citealt{kennicutt1998}) are remarkably well correlated in the inner few kpc radius (Fig.~\ref{multi}). The locations of local CO peaks are generally coincident with the H$\alpha$ knots as shown by high-resolution data. The extents and overall morphology are also similar to each other as seen in the comparison with single-dish data, although the H$\alpha$ blobs outside the old stellar disc do not always have a CO counterpart (e.g. in both ends of NGC~4330's FUV disc; Fig.~\ref{multi}a).

Meanwhile, FUV (a star formation tracer of a time-scale of $\sim$100 Myr; \citealt{kennicutt1998}) is somewhat distinct from CO and H$\alpha$ in extent and morphology, at least on one side, with FUV being more extended than CO and H$\alpha$. Although the FUV disc of many field galaxies are larger than their H$\alpha$ disc (e.g. \citealt{leroy2008}), these three galaxies are distinct from the field population in a sense that both FUV and H$\alpha$ are highly truncated within the stellar disc. These observations indicate that star-forming discs have been shrunk considerably (by a factor of up to 1.5) while undergoing strong ram pressure. Then the question is what causes the quenching of star formation, whether it is molecular gas stripping, the change of molecular gas properties, or both.

As seen by the morphological/kinematical peculiarities of molecular gas in our sample, the ICM pressure is strong enough to strip the H{\sc i} gas in the outer disc or push the H{\sc i} disc to one side, and seems to be able to also affect the molecular gas at a few kpc radii of the stellar disc. This can cause a number of consequences. First, in the outer disc where stars must have been formed mostly from low density interstellar gas, star formation activity will gradually decrease as the atomic gas is stripped. Even when quenching starts, some young stars from the last star formation event may be still observable, and indeed, the FUV emission outside CO and H$\alpha$ discs yet inside the old stellar disc could be good examples of this process.

Our SMA data suggest that these galaxies are close to or approaching peak pressure \citep{kenney2004, crowl2005, chung2007, vollmer2009, abramson2011, vollmer2012a, abramson2014}, which is also supported by the results of simulations in the cases of NGC~4330 and NGC~4522 \citep{vollmer2006,vollmer2009,vollmer2012b}. Therefore, they will experience strong(er) ram pressure for a while, losing more H{\sc i}. As long as H{\sc i} is supplied into the disc, the galaxy should be capable of forming molecular gas, keeping up with star formation. Hence, within one galaxy where star formation can be regulated, the FUV morphology should be more or less in good agreement with those of H$\alpha$, which traces more recent star formation on a time-scale of $<$ 20 Myr \citep{kennicutt1998}. This is indeed true for isolated galaxies such as the sample of \cite{leroy2008}. In addition, the FUV extent in isolated galaxies is usually smaller than the H{\sc i} extent as shown in the comparison of radial profiles \citep{leroy2008}. In our galaxies, however, the H$\alpha$ looks much smaller than the FUV disc, and quite distinct in morphology. This implies that star-forming disc in these galaxies has been shrunken at least in the last 100 Myr.  

\begin{figure*}
\begin{center}
\includegraphics[width=1.0\textwidth,bb=10 330 585 515]{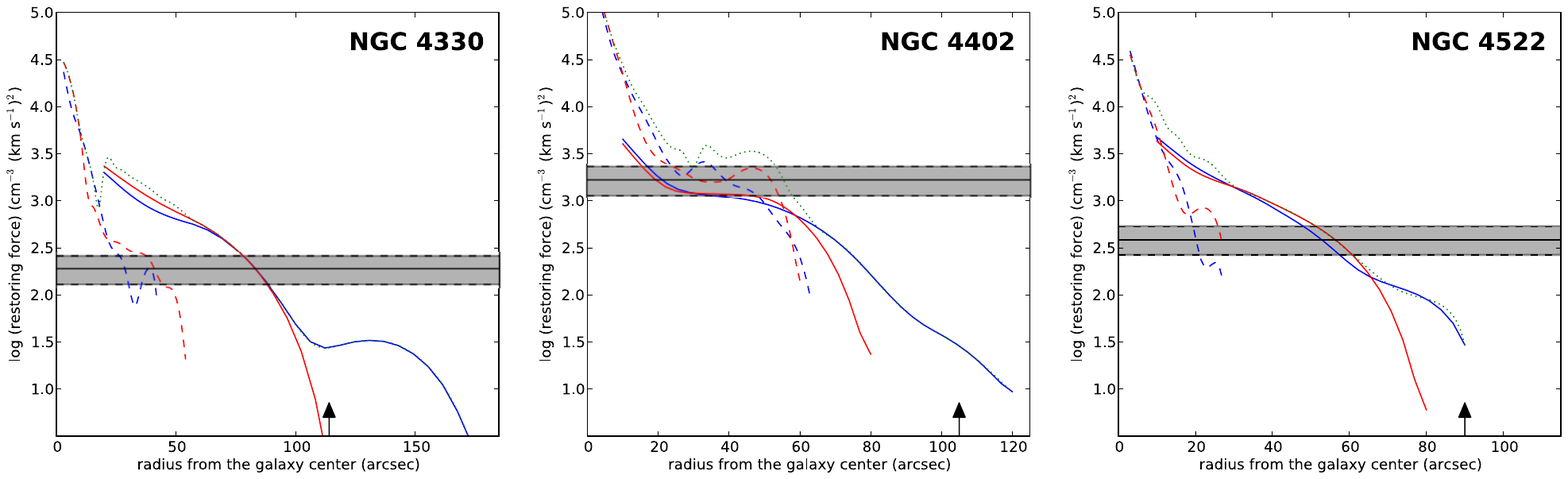}
\caption[]{
The gravitational restoring force acting on the molecular gas of our sample is probed as a function of galactic radius and compared with the current ram pressure estimate. The dashed and the solid lines represent the molecular and the atomic gas, respectively, with the receding side in red and the approaching side in blue. The dotted green line represents the maximum total gas surface density, i.e. $\sum({\rm HI_{max}} + {\rm {H_{2}}_{max}})$ measured at given radii. The upper arrow at the bottom of each panel corresponds to the galaxy radius measured in $R$ band. The vertical range in the ram pressure value (shaded area) corresponds to a range in orbital velocities of each galaxy at a given projected cluster centric distance from the model of \cite{vollmer2001}. Detailed parameters adopted in this plot can be found in Table~\ref{tab4}.
}
 
\label{restore}
\end{center}
\end{figure*}

While the outer disc undergoes star formation truncation, the inner disc is also less likely to form stars in the same manner as field galaxies, despite enough molecular gas present. In the upstream side, molecular gas is found to be compressed as if it is pushed by the ICM pressure. This can be the result of direct molecular gas compression but also molecular gas excess due to formation from compressed diffuse gas \citep{henderson2016}. Consequently, the star formation rate is likely to be locally increased as \cite{fujita1999} suggested (see also \citealt{kronberger2008, bekki2014}). Indeed, we find strong FUV enhancement on the same side where the outer CO disc is found to be compressed. 

Alternatively, this might be the result of the removal of dust due to the ICM pressure and hence the upstream side being unveiled. However, the FUV disc also shows morphological peculiarities and it is quite distinct from the old stellar disc (e.g. a tail in NGC~4330, different extents and scale heights of NGC~4402, extraplanar FUV emission in NGC~4522), which cannot be explained by dust stripping alone. In addition, more evidence for increased star formation due to gas compression by the ICM pressure has been found in other galaxies (e.g. \citealt{merluzzi2013, chung2014}), which also seems to be the case in our sample. Local FUV enhancements found near CO compression within the old stellar disc (such as the north-east of NGC~4330, the east of NGC~4402 and the south-east of NGC~4522) can be good examples of induced star formation by ram pressure (Figs~\ref{n4330}e--\ref{n4522}e and Figs~\ref{multi}a--c).

Some interstellar gas removed from the main stellar disc yet lingering around in a galactic halo may be re-accreted onto the disc after core crossing. NGC~4569 shown in Fig.~\ref{multi}(d), is known to be such case. The single-arm structure in H{\sc i} agrees well in morphology and kinematics with a model of stripped gas falling back to the galaxy, which assumes that the galaxy is currently moving away from the cluster centre after crossing the core $\sim$a few hundreds Myr ago \citep{vollmer2004}. Its star formation quenching time-scale estimated by \cite{crowl2008} is also somewhat longer than our SMA sample by $\sim$100 Myr or so, indicating that this galaxy has experienced strong ICM pressure relatively longer ago than the other three galaxies.

What is intriguing in the case of NGC~4569 is that the extent of the FUV disc is comparable to that of the H$\alpha$ and CO (Fig.~\ref{multi}d). In this case, the star formation is limited in the central region of the galaxy as most of the H{\sc i} gas at larger galactic radii is stripped. Some gas can be re-obtained but not enough gas is left to replenish the entire disc. Therefore, the FUV extent can decrease to a similar size as the H$\alpha$ extent because FUV emission may fade as time goes by, if molecular gas cannot be replenished because the atomic gas has been stripped. At this stage, it is similar to isolated galaxies in a sense that FUV is overall well correlated with H$\alpha$/CO, but now the star-forming disc is somewhat shrunken in comparison to what it used to be during active H{\sc i} stripping or before. The three galaxies that we observed using the SMA are expected to go through a similar evolutionary sequence and will be found with a truncated FUV disc of a comparable size to the H$\alpha$ and CO disc within their old stellar disc. Judging from the star formation quenching timescales by \cite{crowl2008} and simulations by \cite{vollmer2008, vollmer2012a}, the time it will take for our SMA sample to become a galaxy like NGC~4569 is expected to be only a few hundreds Myr.

\begin{table*}
\centering
\begin{minipage}{100mm}
\caption{Parameters of the restoring force and ram pressure.}
\begin{tabular}{lccc}
\hline & NGC~4330 & NGC~4402 & NGC~4522 \\ 
\hline
	& \multicolumn{3}{c}{--- ram pressure parameters ---} \\
$d_{\rmn{M87}} ({\rmn{\degr}}$)$^{\rm a}$ & 2.1 & 1.4 & 3.3 \\
$\rho_{\rm ICM} (10^{-5}~\rmn{cm^{-3}})$$^{\rm b}$ & 8.2 & 63.8 & 20.4 \\
$V_{\rm gal} (\rmn{km~s^{-1}})$$^{\rm c}$ & 1255--1775 & 1338--1892 & 1138--1609\\ 
\noalign{\vspace{-0.3cm}} \\
& \multicolumn{3}{c}{--- restoring force parameters ---} \\
$R$ (kpc)$^{\rm d}$ & 0.2--4.2 & 0.2--4.9 & 0.2--2.1 \\
$\Sigma_{\rm H_{2}} (10^{19}~\rmn{cm^{-2}})$$^{\rm e}$ & 1.4--109 & 6.6--568 & 1.5--187 \\
$V_{\rm rot} (\rmn{km~s^{-1}})$$^{\rm f}$ & 140 & 145 & 122 \\
\hline
\end{tabular}
\label{tab4}

$^{a}${Projected distance from M87.} \\
$^{b}${ICM density.} \\
$^{c}${Range of possible orbital velocities.} \\
$^{d}${Radius from the galactic centre (minimum and maximum values from either side).} \\
$^{e}${Molecular gas surface density (minimum and maximum values from either side).} \\
$^{f}${Rotational velocity, $V_{\rm rot} = \frac {W_{\rm HI,20}} {2~{\rm sin}~i}$, $i$: the inclination angle, $W_{\rm HI,20}$: the H{\sc i} linewidth \citep{chung2009}.} \\
\end{minipage}
\end{table*}

\subsubsection{Extraplanar space}
While a galaxy is undergoing very strong ram pressure, some molecular gas might be pushed away from the stellar disc as clearly seen in NGC~4522. In order to test whether dense molecular gas can get stripped from a galaxy or not, we compare the ram pressure acting on our sample due to the ICM wind with the restoring force (per unit area) by the galactic potential as a function of galactic radii (i.e. $\rho_{\rm ICM}~V_{\rm gal}^2$ versus $V_{\rm rot}^2 R^{-1} \Sigma_{\rm gas}$) in Fig.~\ref{restore}. The ICM density and the orbital velocity of each galaxy are adopted from the model of the Virgo cluster \citep{vollmer2001,yoon2013}. The shaded region indicates the range of possible orbital velocities at a given projected clustercentric distance. To estimate the restoring force at given galactic radii, we take the mean molecular gas and the H{\sc i} gas surface densities along the major axis from the radial profiles (i.e. Figs~\ref{n4330}c--\ref{n4522}c) using our SMA CO data and the H{\sc i} data (VIVA; \citealt{chung2009}). Hence, in this calculation we do not include the IRAM data of NGC~4522, and therefore the restoring force of this case in Fig.~\ref{restore} is only the lower limit of the lowest possible value. The parameters used to estimate the restoring force and ram pressure are summarized in Table~\ref{tab4}.

Our rough estimate suggests that it is not impossible to remove the molecular gas in the very outer region in some circumstances. However, by adopting the rotational velocity of the CO disc, which in all cases has a radius less than 5 kpc, we are focused on a region whose potential is dominated by the stellar component, not the dark matter halo. Therefore, in practice, the actual stripping of molecular gas must be more difficult than our estimation. Also, depending on the encounter angle between the ICM wind and the galactic disc, the inner molecular gas can be better shielded, which will make the stripping even harder. Lastly, considering that the ISM is one entity after all and the total gas surface density of the two phases is more appropriate for more realistic estimation, the chance for the molecular gas to get stripped can be even less. In addition, even if molecular gas is pushed off from the galactic disc, whether the extraplanar molecular gas leaves the galactic halo for good or partially comes back is still debatable. Considering that cluster galaxies are not significantly different from their field counterpart in CO luminosity \citep{stark1986,kenney1989}, the molecular gas completely stripped out of a galaxy is inferred to be a very small fraction.

Our estimate is, however, only the prediction for how easy or hard it is to remove the molecular gas from a galaxy completely, and the molecular gas can be still disturbed without stripping, as our data clearly show. In fact, the recent result by \cite{fumagalli2009} that severely H{\sc i}-deficient galaxies tend to be found with lower molecular gas surface density may simply reflect the fact that the molecular gas of galaxies under strong ICM pressure is perturbed, not necessarily implying molecular gas being deficient.

Extraplanar molecular gas may also form new stars (e.g. ESO~137-001 with a prominent RPS tail found in a massive cluster, Abell~3627; \citealt{sun2007,sun2010,jachym2014}). The FUV emission with several H$\alpha$ knots along the H{\sc i} tail of NGC~4330 and the extraplanar FUV/H$\alpha$ emission in NGC~4522 might be such a case. Outside the old stellar disc, stars may form in two ways, from stripped molecular gas, or alternatively from newly formed molecular gas. Indeed, \cite{tonnesen2012} in their simulations show that stripped diffuse gas can form dense molecular gas by cooling process, which can possibly form new stars. However, in extraplanar space, it is questionable how efficiently stripped atomic hydrogen can be pressurized, turning into molecular form \citep{blitz2006}. Therefore, star formation taking place outside the main disc is more likely to be utilizing the molecular gas that is pushed from the main disc but has not left the galactic halo.

As in NGC~4522, however, there are several extraplanar H$\alpha$ regions with no detectable CO. This suggests that the H$_{2}$ gas depletion time-scale ($\tau_{\rm H_2}$) in extraplanar space can be much shorter than what it is on the disc (of the order of a few Gyr; \citealt{bigiel2008,bigiel2011}). Meanwhile, $\tau_{\rm H_2}$ at the location of H$\alpha$/CO blob is measured to be comparable to what is normally found in the galactic disc (see Appendix B). Therefore, there might be some extraplanar CO even in NGC~4330 and NGC~4402, but with mass smaller than our detectability or that of the IRAM \citep{vollmer2008}. 

Together with this, there is also an important difference between NGC~4330 and NGC~4522, which is the presence of CO in the extraplanar star formation region. In the case of NGC~4330, more rapidly increasing ram pressure compared to NGC~4522 might be responsible for faster dissociation of molecular cloud, and hence absence of CO along the FUV tail. Alternatively, some fraction of stripped molecular gas might be destroyed by high-energy photon emitted from massive young stars \citep{crowl2005}. Therefore extraplanar or stripped CO may not be always observable even if it is pushed off from the stellar disc.

\section{CONCLUSIONS}
\label{sec6}
In this work, we have presented high-resolution 12CO (2$-$1) data obtained using the SMA of three Virgo spirals, NGC~4330, NGC~4402 and NGC~4522, which are undergoing active H{\sc i} stripping due to ram pressure. By comparing with single-dish 12CO (2$-$1), H{\sc i}, FUV, H$\alpha$ and DSS2 red data, we have discussed how ICM pressure changes molecular gas and star formation properties. Most important results from our study can be summarized as follows: 

\begin{enumerate}
\item The overall CO morphology and kinematics are quite asymmetric and disturbed. Morphological peculiarities found in CO are closely related to H{\sc i} morphology in these galaxies, indicating that molecular gas is also affected by the ram pressure.

\item We find the CO clump in the south-west of NGC~4330 (Figs~\ref{n4330}a--c) and the modest CO bump in the north-west of NGC~4402 (Figs~\ref{n4402}a--c), with distinct velocity structure from the overall velocity gradient of the molecular disc. This also supports the idea that molecular gas can be displaced by the external pressure inside a stellar disc.

\item Both FUV and H$\alpha$ are enhanced where H{\sc i} and CO are compressed, supporting that ram pressure can also trigger star formation temporarily (e.g. Figs~\ref{n4402}d--f of NGC~4402). In these galaxies, however, FUV shows distinct morphology and extent from those of CO, while H$\alpha$ and CO are overall in good agreement (Fig.~\ref{multi}). The large offset between FUV and H$\alpha$ must indicate the star formation has been recently quenched, likely over the last 100 Myr as the molecular gas properties have changed.

\item To compare with NGC~4569 (Fig.~\ref{multi}d), which is thought to have already crossed the cluster core, the offset between FUV and H$\alpha$/CO is expected to smooth out over time ($\sim$a few hundreds Myr) after the core crossing as the galaxy further loses H{\sc i}, i.e. the capability of forming molecular gas, and star formation is completely shut down at large galactic radii.
\end{enumerate}

In summary, our results suggest that a strong ICM pressure not only strips diffuse atomic gas but also changes the properties of dense molecular gas in the inner few kpc of a galaxy. Molecular gas can be pushed from the stellar disc in the outer part as reported in previous studies, yet we find that molecular gas is less likely to be completely stripped from a galaxy. On the side where the ICM pressure acts, molecular gas can be enhanced by compression and molecular formation from atomic gas, locally increasing star formation. As the gas is rapidly used up, the star-forming disc will decrease in size as no sufficient atomic gas is left to form more molecular gas. On the opposite side, molecular gas can be perturbed in a way that reflects the atomic gas, which may prevent a galaxy from forming stars as before, i.e. injecting turbulence into molecular gas is likely to cause star formation rates to be inefficient \citep{alatalo2015}. The combination of these two effects must be an important driver that makes galaxies evolve to become passive in the cluster environment.   

In order to investigate the molecular gas properties and its conditions for star formation under strong ICM pressure more quantitatively, we are currently probing the density and temperature of molecular gas using multitransitions and isotopes of CO (Lee et al. in preparation). Our future work is expected to provide further insights on the evolution of molecular gas and thus star formation efficiency in galaxies undergoing ram pressure.

\section*{ACKNOWLEDGEMENTS}
Support for this work was provided by the National Research Foundation of Korea to the Center for Galaxy Evolution Research (No. 2010-0027910) and Science Fellowship of POSCO TJ Park Foundation. This work has been also supported by NRF grant No. 2015R1D1A1A01060516. The Submillimeter Array is a joint project between the Smithsonian Astrophysical Observatory and the Academia Sinica Institute of Astronomy and Astrophysics and is funded by the Smithsonian Institution and the Academia Sinica. This work made use of HERACLES, `The HERA CO-Line Extragalactic Survey' \citep{leroy2009}. We acknowledge the usage of the HyperLeda database (http://leda.univ-lyon1.fr). This research has made use of the GOLDMine Database \citep{gavazzi2003}, and it made use of \textsc{aplpy}, an open-source plotting package for \textsc{python} hosted at http://aplpy.github.com.  ST was supported by the Alvin E. Nashman Fellowship in Theoretical Astrophysics. BL is grateful to E.-J. Chung for the technical help with \textsc{gipsy}.





\begin{thebibliography}{99}
\bibitem[\protect\citeauthoryear{Abramson et al.}{2011}]{abramson2011} Abramson A., Kenney J.~D.~P., Crowl H.~H., Chung A., van Gorkom J.~H., Vollmer B., Schiminovich D., 2011, AJ, 141, 164
\bibitem[\protect\citeauthoryear{Abramson \& Kenney}{2014}]{abramson2014} Abramson A., Kenney J.~D.~P., 2014, AJ, 147, 63 
\bibitem[\protect\citeauthoryear{Alatalo et al.}{2015}]{alatalo2015} Alatalo K., et al., 2015, ApJ, 812, 117 


\bibitem[\protect\citeauthoryear{Barmby et al.}{2006}]{barmby2006} Barmby P., et al., 2006, ApJ, 650, L45
\bibitem[\protect\citeauthoryear{Bekki}{2014}]{bekki2014} Bekki K., 2014, MNRAS, 438, 444
\bibitem[\protect\citeauthoryear{Bigiel et al.}{2008}]{bigiel2008} Bigiel F., Leroy A., Walter F., Brinks E., de Blok W.~J.~G., Madore B., Thornley M.~D., 2008, AJ, 136, 2846
\bibitem[\protect\citeauthoryear{Bigiel et al.}{2011}]{bigiel2011} Bigiel F., et al., 2011, ApJ, 730, L13 
\bibitem[\protect\citeauthoryear{Blitz \& Rosolowsky}{2006}]{blitz2006} Blitz L., Rosolowsky E., 2006, ApJ, 650, 933
\bibitem[\protect\citeauthoryear{B{\"o}hringer et al.}{1994}]{bohringer1994} B{\"o}hringer H., Briel U.~G., Schwarz R.~A., Voges W., Hartner G., Tr{\"u}mper J., 1994, Nature, 368, 828 
\bibitem[\protect\citeauthoryear{Boselli et al.}{2014}]{boselli2014} Boselli A., Cortese L., Boquien M., Boissier S., Catinella B., Gavazzi G., Lagos C., Saintonge A., 2014, A\&A, 564, A67 


\bibitem[\protect\citeauthoryear{Cayatte et al.}{1990}]{cayatte1990} Cayatte V., van Gorkom J.~H., Balkowski C., Kotanyi C., 1990, AJ, 100, 604 
\bibitem[\protect\citeauthoryear{Chung et al.}{2007}]{chung2007} Chung A., van Gorkom J.~H., Kenney J.~D.~P., Vollmer B., 2007, ApJ, 659, L115
\bibitem[\protect\citeauthoryear{Chung et al.}{2009}]{chung2009} Chung A., van Gorkom J.~H., Kenney J.~D.~P., Crowl H., Vollmer B., 2009, AJ, 138, 1741 
\bibitem[\protect\citeauthoryear{Chung}{2012}]{chung2012} Chung E.~J., 2012, PhD thesis, Yonsei University
\bibitem[\protect\citeauthoryear{Chung \& Kim}{2014}]{chung2014} Chung E.~J., Kim S., 2014, PASJ, 66, 11
\bibitem[\protect\citeauthoryear{Crowl et al.}{2005}]{crowl2005} Crowl H.~H., Kenney J.~D.~P., van Gorkom J.~H., Vollmer B., 2005, AJ, 130, 65 
\bibitem[\protect\citeauthoryear{Crowl \& Kenney}{2008}]{crowl2008} Crowl H.~H., Kenney J.~D.~P., 2008, AJ, 136, 1623


\bibitem[\protect\citeauthoryear{Davies \& Lewis}{1973}]{davies1973} Davies R.~D., Lewis B.~M., 1973, MNRAS, 165, 231
\bibitem[\protect\citeauthoryear{Drimmel \& Spergel}{2001}]{drimmel2001} Drimmel R., Spergel D.~N., 2001, ApJ, 556, 181

\bibitem[\protect\citeauthoryear{Fujita \& Nagashima}{1999}]{fujita1999} Fujita Y., Nagashima M., 1999, ApJ, 516, 619 
\bibitem[\protect\citeauthoryear{Fumagalli et al.}{2009}]{fumagalli2009} Fumagalli M., Krumholz M.~R., Prochaska J.~X., Gavazzi G., Boselli A., 2009, ApJ, 697, 1811


\bibitem[\protect\citeauthoryear{Gavazzi et al.}{2003}]{gavazzi2003} Gavazzi G., Boselli A., Donati A., Franzetti P., Scodeggio M., 2003, A\&A, 400, 451 
\bibitem[\protect\citeauthoryear{Gil de Paz et al.}{2007}]{gil2007} Gil de Paz A., et al., 2007, ApJS, 173, 185
\bibitem[\protect\citeauthoryear{Giovanelli \& Haynes}{1985}]{giovanelli1985} Giovanelli R., Haynes M.~P., 1985, ApJ, 292, 404
\bibitem[\protect\citeauthoryear{Gunn \& Gott}{1972}]{gunn1972} Gunn J.~E., Gott J.~R., III, 1972, ApJ, 176, 1

\bibitem[\protect\citeauthoryear{Haynes \& Giovanelli}{1984}]{haynes1984} Haynes M.~P., Giovanelli R., 1984, AJ, 89, 758
\bibitem[\protect\citeauthoryear{Helfer et al.}{2003}]{helfer2003} Helfer T.~T., Thornley, M.~D., Regan, M.~W., Wong T., Sheth K., Vogel S.~N., Blitz L., Bock D.~C.-J., 2003, ApJS, 145, 259 
\bibitem[\protect\citeauthoryear{Henderson \& Bekki}{2016}]{henderson2016} Henderson B., Bekki K., 2016, ApJ, 822, L33 
\bibitem[\protect\citeauthoryear{Hidaka \& Sofue}{2002}]{hidaka2002} Hidaka M., Sofue Y., 2002, PASJ, 54, 33 


\bibitem[\protect\citeauthoryear{J{\'a}chym et al.}{2014}]{jachym2014} J{\'a}chym P., Combes F., Cortese L., Sun M., Kenney J.~D.~P., 2014, ApJ, 792, 11 

\bibitem[\protect\citeauthoryear{Kenney \& Young}{1989}]{kenney1989} Kenney J.~D.~P., Young J.~S., 1989, ApJ, 344, 171 
\bibitem[\protect\citeauthoryear{Kenney et al.}{1990}]{kenney1990} Kenney J.~D.~P., Young J.~S., Hasegawa T., Nakai N., 1990, ApJ, 353, 460
\bibitem[\protect\citeauthoryear{Kenney et al.}{2004}]{kenney2004} Kenney J.~D.~P., van Gorkom J.~H., Vollmer B., 2004, AJ, 127, 3361 
\bibitem[\protect\citeauthoryear{Kenney et al.}{2015}]{kenney2015} Kenney J.~D.~P., Abramson A., Bravo-Alfaro H., 2015, AJ, 150, 59 
\bibitem[\protect\citeauthoryear{Kennicutt}{1998}]{kennicutt1998} Kennicutt R.~C., Jr., 1998, ARA\&A, 36, 189
\bibitem[\protect\citeauthoryear{Koopmann et al.}{2001}]{koopmann2001} Koopmann R.~A., Kenney J.~D.~P., Young J., 2001, ApJS, 135, 125 
\bibitem[\protect\citeauthoryear{Koopmann \& Kenney}{2004a}]{koopmann2004a} Koopmann R.~A., Kenney J.~D.~P., 2004a, ApJ, 613, 851
\bibitem[\protect\citeauthoryear{Koopmann \& Kenney}{2004b}]{koopmann2004b} Koopmann R.~A., Kenney J.~D.~P., 2004b, ApJ, 613, 866
\bibitem[\protect\citeauthoryear{Kregel, van der Kruit \& de Blok}{2004}]{kregel2004} Kregel M., van der Kruit P.~C., de Blok W.~J.~G., 2004, MNRAS, 352, 768 
\bibitem[\protect\citeauthoryear{Kronberger et al.}{2008}]{kronberger2008} Kronberger T., Kapferer W., Ferrari C., Unterguggenberger S., Schindler S., 2008, A\&A, 481, 337 



\bibitem[\protect\citeauthoryear{Lee \& Chung}{2015}]{lee2015} Lee B., Chung A., 2015, PKAS, 30, 491 
\bibitem[\protect\citeauthoryear{Leroy et al.}{2008}]{leroy2008} Leroy A.~K., Walter F., Brinks E., Bigiel F., de Blok W.~J.~G., Madore B., Thornley M.~D., 2008, AJ, 136, 2782
\bibitem[\protect\citeauthoryear{Leroy et al.}{2009}]{leroy2009} Leroy A.~K., et al., 2009, AJ, 137, 4670 
\bibitem[\protect\citeauthoryear{Lucy}{1974}]{lucy1974} Lucy L.~B., 1974, AJ, 79, 745

\bibitem[\protect\citeauthoryear{Mei et al.}{2007}]{mei2007} Mei S., et al., 2007, ApJ, 655, 144 
\bibitem[\protect\citeauthoryear{Merluzzi et al.}{2013}]{merluzzi2013} Merluzzi P., et al., 2013, MNRAS, 429, 1747 
\bibitem[\protect\citeauthoryear{Mok et al.}{2016}]{mok2016} Mok A., et al., 2016, MNRAS, 456, 4384 

\bibitem[\protect\citeauthoryear{Nehlig et al.}{2016}]{nehlig2016} Nehlig F., Vollmer B., Braine J., 2016, A\&A, 587, A108 

\bibitem[\protect\citeauthoryear{Paturel et al.}{2003}]{paturel2003} Paturel G., Petit C., Prugniel P., Theureau G., Rousseau J., Brouty M., Dubois P., Cambr{\'e}sy L., 2003, A\&A, 412, 45 

\bibitem[\protect\citeauthoryear{Qi}{2012}]{qi2012} Qi C., 2012, Available at: http://www.cfa.harvard.edu/$\sim$cqi/mircook.html

\bibitem[\protect\citeauthoryear{Rengarajan \& Iyengar}{1992}]{rengarajan1992} Rengarajan T.~N., Iyengar K.~V.~K., 1992, MNRAS, 259, 559
\bibitem[\protect\citeauthoryear{Rhee \& van Albada}{1996}]{rhee1996} Rhee M.-H., van Albada T.~S., 1996, A\&AS, 115, 407


\bibitem[\protect\citeauthoryear{Salo et al.}{2015}]{salo2015} Salo H., et al., 2015, ApJS, 219, 4
\bibitem[\protect\citeauthoryear{Skrutskie et al.}{2006}]{skrutskie2006} Skrutskie M.~F., et al., 2006, AJ, 131, 1163 
\bibitem[\protect\citeauthoryear{Sofue et al.}{2003a}]{sofue2003a} Sofue Y., Koda J., Nakanishi H., Onodera S., Kohno K., Tomita A., Okumura S.~K.,  2003a, PASJ, 55, 17 
\bibitem[\protect\citeauthoryear{Sofue et al.}{2003b}]{sofue2003b} Sofue Y., Koda J., Nakanishi H., Onodera S., 2003b, PASJ, 55, 59 
\bibitem[\protect\citeauthoryear{Sofue et al.}{2003c}]{sofue2003c} Sofue Y., Koda J., Nakanishi H., Hidaka M., 2003c, PASJ, 55, 75 
\bibitem[\protect\citeauthoryear{Solomon et al.}{1997}]{solomon1997} Solomon P.~M., Downes D., Radford S.~J.~E., Barrett J.~W., 1997, ApJ, 478, 144 
\bibitem[\protect\citeauthoryear{Stark et al.}{1986}]{stark1986} Stark A.~A., Knapp G.~R., Bally J., Wilson R.~W., Penzias A.~A., Rowe H.~E., 1986, ApJ, 310, 660
\bibitem[\protect\citeauthoryear{Strong \& Mattox}{1996}]{strong1996} Strong A.~W., Mattox J.~R., 1996, A\&A, 308, L21
\bibitem[\protect\citeauthoryear{Sun, Donahue \& Voit}{2007}]{sun2007} Sun M., Donahue M., Voit G.~M., 2007, ApJ, 671, 190  
\bibitem[\protect\citeauthoryear{Sun et al.}{2010}]{sun2010} Sun M., Donahue M., Roediger E., Nulsen P.~E.~J., Voit G.~M., Sarazin C., Forman W., Jones C., 2010, ApJ, 708, 946 

\bibitem[\protect\citeauthoryear{Swaters}{1999}]{swaters1999} Swaters R.~A., 1999, PhD thesis, University of Groningen  

\bibitem[\protect\citeauthoryear{Tonnesen \& Bryan}{2012}]{tonnesen2012} Tonnesen S., Bryan G.~L., 2012, MNRAS, 422, 1609

\bibitem[\protect\citeauthoryear{Verheijen}{1997}]{verheijen1997}Verheijen M.~A.~W., 1997, PhD thesis, University of Groningen
\bibitem[\protect\citeauthoryear{Vollmer et al.}{2001}]{vollmer2001} Vollmer B., Cayatte V., Balkowski C., Duschl W.~J., 2001, ApJ, 561, 708 

\bibitem[\protect\citeauthoryear{Vollmer et al.}{2004}]{vollmer2004} Vollmer B., Balkowski C., Cayatte V., van Driel W., Huchtmeier W., 2004, A\&A, 419, 35
\bibitem[\protect\citeauthoryear{Vollmer et al.}{2006}]{vollmer2006} Vollmer B., Soida M., Otmianowska-Mazur K., Kenney J.~D.~P., van Gorkom J.~H., Beck R., 2006, A\&A, 453, 883 
\bibitem[\protect\citeauthoryear{Vollmer et al.}{2008}]{vollmer2008} Vollmer B., Braine J., Pappalardo C., Hily-Blant P., 2008, A\&A, 491, 455

\bibitem[\protect\citeauthoryear{Vollmer}{2009}]{vollmer2009} Vollmer B., 2009, A\&A, 502, 427 
\bibitem[\protect\citeauthoryear{Vollmer et al.}{2012a}]{vollmer2012a} Vollmer B., et al., 2012a,  A\&A, 537, A143 
\bibitem[\protect\citeauthoryear{Vollmer et al.}{2012b}]{vollmer2012b} Vollmer B., Wong O.~I., Braine J., Chung A., Kenney J.~D.~P., 2012b, A\&A, 543, A33 


\bibitem[\protect\citeauthoryear{Walter et al.}{2008}]{walter2008} Walter F., Brinks E., de Blok W.~J.~G., Bigiel F., Kennicutt R.~C., Jr., Thornley M.~D., Leroy A., 2008, AJ, 136, 2563
\bibitem[\protect\citeauthoryear{Warmels}{1988a}]{warmels1988a} Warmels R.~H., 1988a, A\&AS, 72, 19
\bibitem[\protect\citeauthoryear{Warmels}{1988b}]{warmels1988b} Warmels R.~H., 1988b, A\&AS, 72, 427 
\bibitem[\protect\citeauthoryear{Wong et al.}{2014}]{wong2014} Wong O.~I., Kenney J.~D.~P., Murphy E.~J., Helou G., 2014, ApJ, 783, 109 

\bibitem[\protect\citeauthoryear{Yasuda et al.}{1997}]{yasuda1997} Yasuda N., Fukugita M., Okamura S., 1997, ApJS, 108, 417 
\bibitem[\protect\citeauthoryear{Yoon}{2013}]{yoon2013} Yoon H., 2013, MS thesis, Yonsei University
\end{thebibliography}




\appendix

\section{12CO gas distribution on the optical colour image}

\begin{figure*}
\begin{center}
\includegraphics[width=0.9\textwidth,bb=0 90 595 755]{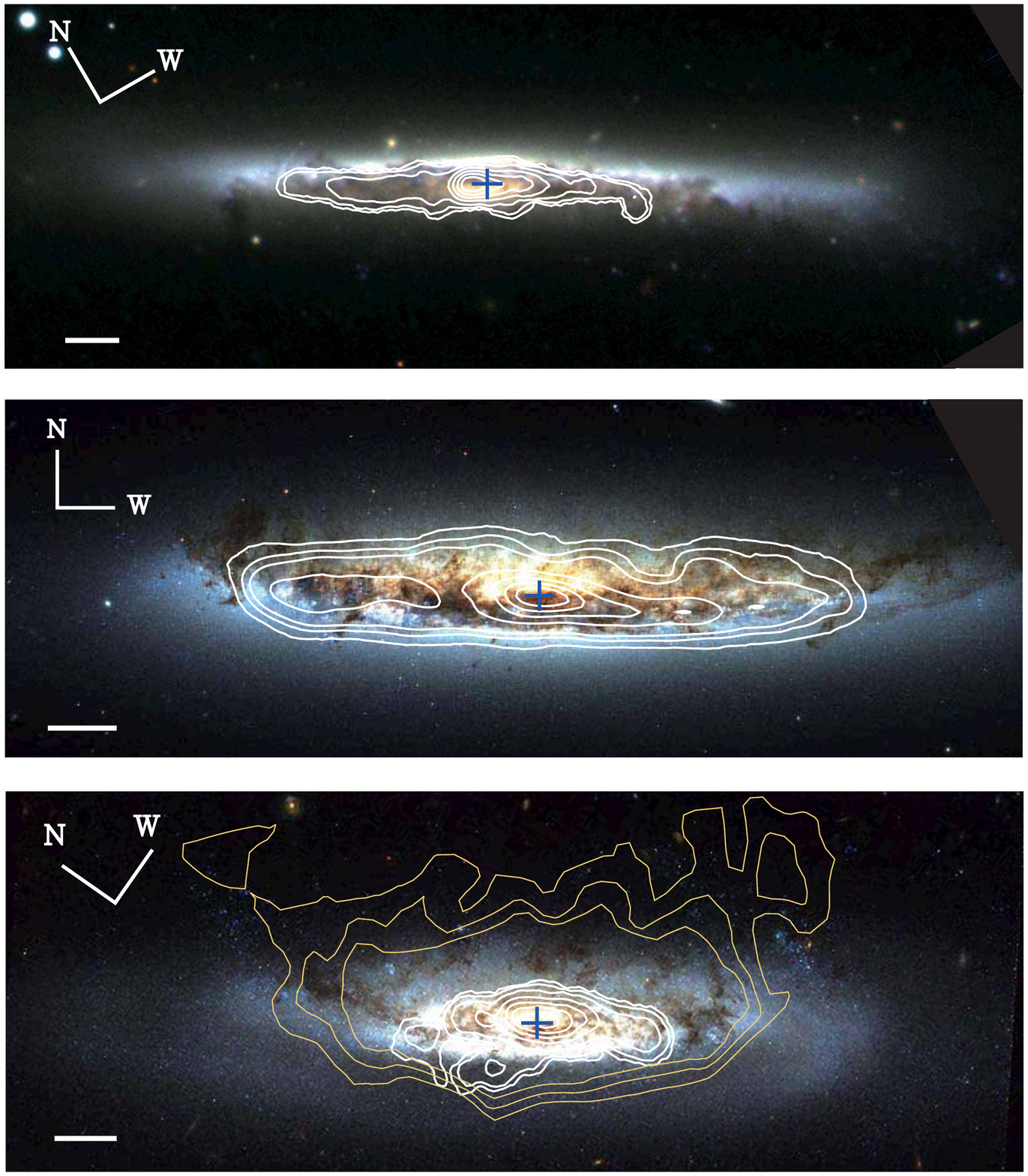}
\caption[]{
12CO (2$-$1) contours (white conours: SMA, yellow contours: IRAM) are overlaid on optical colour images (NGC~4330: WIYN 3.5m telescope BVR colour image; \citealt{abramson2011}, NGC~4402 and NGC~4522: {\it HST}\footnotemark[3] BVI colour images). Top: NGC~4330. Middle: NGC~4402. Bottom: NGC~4522. The physical scale bar (20 arcsec) of each galaxy is shown at the bottom left. The blue cross indicates the stellar disc centre of each galaxy.
}

\label{hst}
\end{center}
\end{figure*}
\stepcounter{footnote}\footnotetext{Based on observations made with the NASA/ESA {\it Hubble Space Telescope}, obtained from the data archive at the Space Telescope Science Institute. STScI is operated by the Association of Universities for Research in Astronomy, Inc. under NASA contract NAS 5-26555.}

\section{The H$_{2}$ gas depletion time-scale ($\tau_{\rm H_2}$) in the extraplanar region of NGC~4522}
\begin{figure*}
\begin{center}
\includegraphics[width=0.9\textwidth,bb=5 150 580 700]{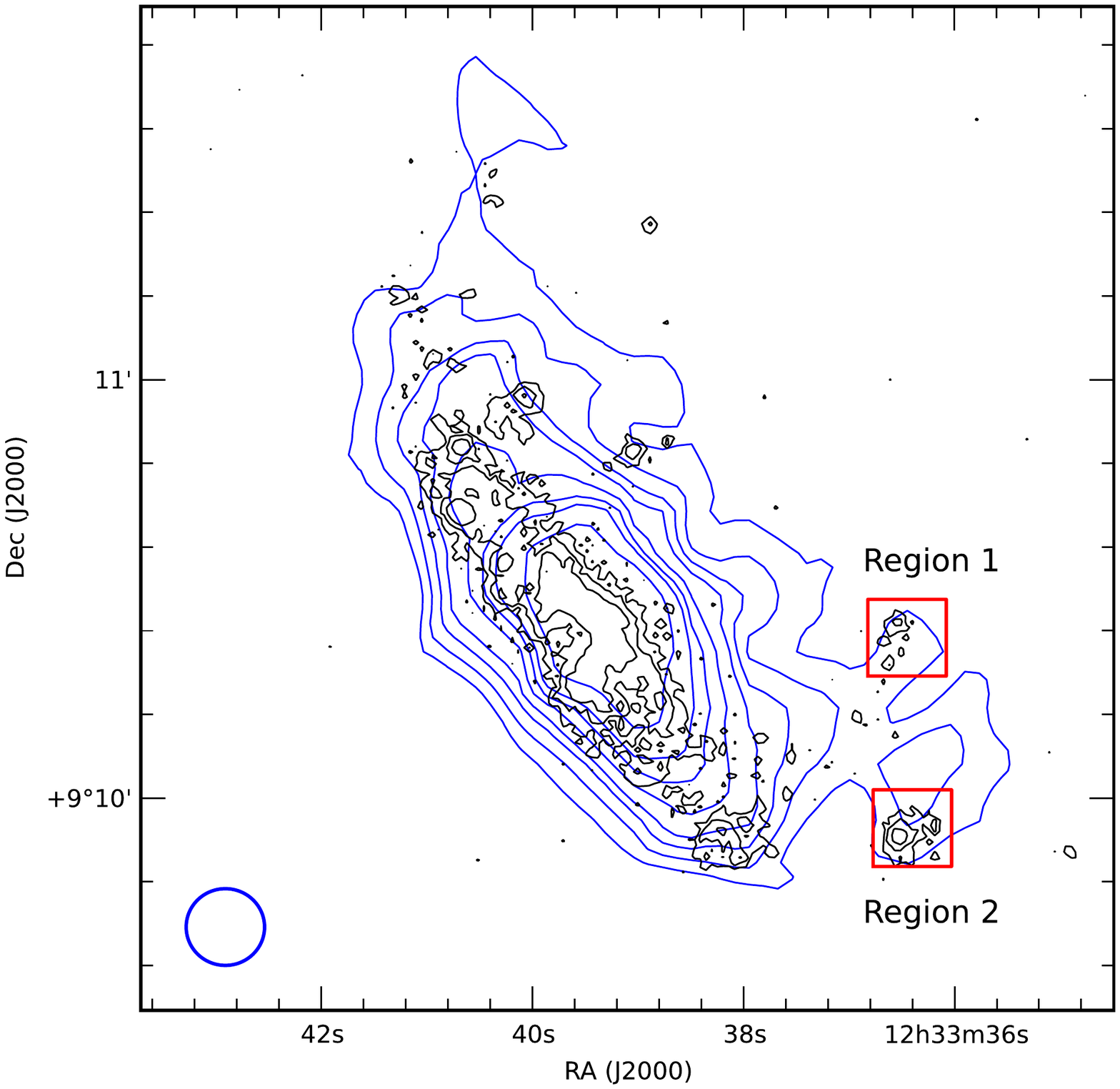}
\caption[]{The H$\alpha$ image is shown in black contours overlaid with IRAM 12CO (2$-$1) image in blue contours. The beam size of the IRAM is shown in blue circle at the bottom left. Two regions with extraplanar H$\alpha$ and CO emissions where the molecular gas depletion time has been estimated are indicated in red boxes.
}

\label{deple}
\end{center}
\end{figure*}

We estimate the molecular gas depletion time-scale (the ratio of molecular gas mass and star formation rate) in extraplanar region of NGC~4522 following the prescription below. We first calculate molecular gas mass in two regions (Figure~\ref{deple}, red squares) using IRAM 12CO (2$-$1) data of \cite{vollmer2008}. Using H$\alpha$ image \citep{koopmann2001}, the star formation rate in each region (Fig.~\ref{deple}, red squares) is estimated as follows:
\begin{equation}
SFR = 7.9 \times 10^{-42}~L({\rm H\alpha})
\end{equation}
in ${\rm M_{\sun}}$~yr$^{-1}$ where $L({\rm H\alpha})$ is the H$\alpha$ luminosity in erg~s$^{-1}$ \citep{kennicutt1998}. Measured depletion time-scales are 1.1 Gyr (region 1) and 0.8 Gyr (region 2), respectively.

\bsp	
\label{lastpage}
\end{document}